\documentclass[a4paper,12pt]{article}
\pdfoutput=1 

\usepackage{amsmath,amssymb}
\usepackage[pdftex]{graphicx}
\usepackage[pdftex]{hyperref}
\usepackage{epsf}
\usepackage{cite}
\usepackage{mathrsfs}
\usepackage[usenames,dvipsnames]{xcolor}
\usepackage{graphicx}

\makeatletter

\newcommand{\dtt}{\widetilde{dt\,}}
\newcommand{\av}{{\mathbf a}}

\newcommand{\xv}{{\mathbf x}}

\newcommand{\Fv}{{\mathbf F}}
\newcommand{\Rv}{{\mathbf R}}

\newcommand{\Lap}{{\bigtriangleup}}

\providecommand{\abs}[1]{\lvert#1\rvert}

\g@addto@macro\bfseries{\boldmath}

\def\refchecklabelfontsize{\fontsize{5pt}{5pt}\selectfont}
\let\mark@size=\refchecklabelfontsize

\def\unit{{1\kern-.65ex {\rm l}}}
\def\1{{1\kern-.65ex {\rm l}}}

\def\bracket#1{{\langle{#1}\rangle}}

\def\cA{{\cal A}}

\def\cF{{\cal F}}

\def\cN{{\cal N}}
\def\cO{{\cal O}}

\def\cR{{\cal R}}

\def\bbR{{\mathbb{R}}}

\def\bbZ{{\mathbb{Z}}}

\newcount\hour \newcount\minute
\hour=\time \divide \hour by 60
\minute=\time
\def\now{%
\ifnum \hour<13
  \ifnum \hour=0 \advance \hour by 12 \number\hour:\else \number\hour:\fi%
     \ifnum \minute<10 0\fi%
     \number\minute%
\ A.M.%
\else \advance \hour by -12 \number\hour:%
  \ifnum \minute<10 0\fi%
  \number\minute%
  \ P.M.%
\fi%
}

\makeatother

\begin{document}


\baselineskip=18pt  
\numberwithin{equation}{section}  







\thispagestyle{empty}

\vspace*{-2cm} 
\begin{flushright}
YITP-15-41
\end{flushright}

\vspace*{2.5cm} 
\begin{center}
 {\LARGE Codimension-2  Solutions \\[1ex] in
 Five-Dimensional Supergravity}\\
 \vspace*{1.7cm}
 Minkyu Park$^1$ and Masaki Shigemori$^{1,2}$\\
 \vspace*{1.0cm} 
 $^1$ Yukawa Institute for Theoretical Physics, Kyoto University\\
Kitashirakawa Oiwakecho, Sakyo-ku, Kyoto 606-8502 Japan\\[1ex]
 $^2$ 
Hakubi Center, Kyoto University\\
Yoshida-Ushinomiya-cho, Sakyo-ku, Kyoto 606-8501, Japan
\end{center}
\vspace*{1.5cm}

\noindent

We study a new class of supersymmetric solutions in five-dimensional
supergravity representing multi-center configurations of codimension-2
branes along arbitrary curves.
Co\-di\-men\-sion-2 
branes are produced in generic situations out of ordinary branes of
higher codimension by the supertube effect and, when they are exotic
branes, spacetime generally becomes non-geometric.  The solutions are
characterized by a set of harmonic functions on $\mathbb{R}^3$ with
non-trivial monodromies around codimension-2 branch-point
singularities.  The solutions can be regarded as generalizations of the
Bates-Denef/Bena-Warner multi-center solutions with codimension-3
centers to include codimension-2 ones.  We present some explicit
examples of solutions with codimension-2 centers, and discuss their
relevance for the black microstate (non-)geometry program.

\newpage
\setcounter{page}{1} 





\section{Introduction}

String theory contains various branes that come in diverse dimensions,
such as D-branes, and they have played a crucial role in understanding
the non-perturbative aspects of string theory.  Among these branes, ones
with small $(\le 2)$ codimension have been relatively less studied,
probably due to their non-standard features. For instance, the
codimension-2 D7-brane destroys the spacetime asymptotics by introducing
conical deficit, and the codimension-1 D8-brane terminates spacetime a
finite distance from it as the dilaton diverges.
However, it is such peculiarities that make small-codimension branes
special and all the more interesting.  For example, the fact that
7-branes change spacetime asymptotics is precisely what makes the
F-theory geometries work \cite{Greene:1989ya, Vafa:1996xn}.  More
recently, it was pointed out \cite{deBoer:2010ud, deBoer:2012ma} that
small-codimension branes can be spontaneously created out of ordinary
(codimension $>2$) branes by the supertube transition
\cite{Mateos:2001qs} and generically lead to non-geometric spacetime.
In particular, black holes in string theory are typically constructed by
intersecting multiple stacks of branes, which can spontaneously polarize
by the supertube transition into small-codimension branes.  So, studying
small-codimension branes and the accompanying non-geometric structure of
spacetime is relevant for understanding microscopic physics of black
holes in string theory.

Five-dimensional supergravity has been extensively used as a convenient
paradigm in which to study black holes in string theory.  In particular,
all supersymmetric solutions of $d=5$, $\cN=1$ ungauged supergravity
with vector multiplets have been completely classified in
\cite{Gauntlett:2002nw, Bena:2004de}.\footnote{For supersymmetric
solutions in more general $d=5$, $\cN=1$ supergravities, such as gauged
theories and theories with hyper and tensor multiplets, see
\cite{Gutowski:2004yv,Gauntlett:2004qy,Gutowski:2005id,Bellorin:2006yr,Bellorin:2007yp,Bellorin:2008we}. } 
This supergravity theory describes the low-energy physics of M-theory compactified on a Calabi-Yau 3-fold
$X$ or, in the presence of an additional $S^1$ \cite{Gauntlett:2004qy,
Gauntlett:2002nw}, of type IIA string theory compactified on $X$.  In the
latter picture, these supersymmetric solutions represent a system of D6,
D4, D2, and D0-branes wrapped on various cycles inside 
$X$ \cite{Bates:2003vx}.  Let us call this solution of $d=5$
supergravity the ``4D/5D solution.''
The 4D/5D solution is completely specified by a set of harmonic
functions, which we collectively denote by $H$, on a spatial $\bbR^3$ 
base. Its general form is
\begin{equation}
 H(\xv)=h+\sum_{p=1}^N {\Gamma_p\over |\xv-\av_p|}\,,\label{Hharmonic}
\end{equation}
and the associated 4D/5D solution represents a bound state of $N$ black
hole centers, which are sitting at $\xv=\av_p$ ($p=1,\dots,N$) and are made of D6, D4, D2, and
D0-branes represented by the coefficients $\Gamma_p$.  The black hole
centers are of codimension 3, being a point in the $\bbR^3$.
%
The 4D/5D solution has been applied to various studies of black holes
and rings in four and five dimensions, such as the black hole
attractor mechanism \cite{Ferrara:1995ih,Strominger:1996kf,Ferrara:1996dd,Ferrara:1996um,Moore:2004fg,Kraus:2005gh,Larsen:2006xm}, split attractor flows and wall crossing
\cite{Denef:2000ar,Denef:2001ix,Moore2010pitp,Denef:2007vg,Bates:2003vx}, and microstate
geometries \cite{Bena:2005va, Berglund:2005vb}.

The supertube transition \cite{Mateos:2001qs} is a spontaneous
polarization phenomenon in which a particular combination of branes
puffs up into a new dipole charge.  For example, if we put two
orthogonal D2-branes together, they will polarize into an NS5-brane
along an arbitrary closed curve parametrized by $\lambda$.  We represent
this process as follows:
\begin{align}
 \rm D2(45)+D2(67)\to ns5(\lambda 4567)\,,
\label{D2D2ns5intro}
\end{align}
where D2(45) denotes the D2-brane wrapped around 45 directions and
``ns5'' in lowercase means that it is a dipole charge.  We assume that
4567 directions are compact.\footnote{Note that the process
\eqref{D2D2ns5intro} is what \emph{will} happen if we put together two
D2-branes preserving supersymmetry.  There is \emph{no} option for them
not to puff up.  Two D2-branes on top of each other, un-puffed up, are
not supersymmetric, unless 4567 directions are non-compact (and thus
branes are infinite in extent) or $g_s=0$; see
\cite[Sec.~3.1]{Mathur:2005zp}.}
As we have mentioned, such D2-branes appear in the 4D/5D solution
described by \eqref{Hharmonic}, and the supertube transition
\eqref{D2D2ns5intro} implies that the solution must actually be extended
to include codimension-2 sources along arbitrary curves in the $\bbR^3$,
in order to describe the full configuration space of the brane system.

As we will see in explicit examples later, this does not just mean to
smear the \mbox{codimension-3} singularities in the harmonic function
\eqref{Hharmonic} along a curve to get a codimension-2 singularity, but
the harmonic function can also have branch-point singularities and be
\emph{multi-valued} in $\bbR^3$.
It is a generic feature of codimension-2 branes that, as one goes around
their worldvolume, the spacetime fields undergo a U-duality
transformation \cite{deBoer:2010ud, deBoer:2012ma} and become
multi-valued; the harmonic function being multi-valued is the
manifestation of this.

For the transition \eqref{D2D2ns5intro}, it is only the $B$-field that
are multi-valued around the supertube (ns5).  However, there are also
supertube transitions that produce non-geometric exotic branes,
around which the metric is multi-valued.  One example is
\begin{equation}
\rm D2(89)+D6(456789)\to 5^2_2(\lambda 4567;89)\,,\label{D2D6522intro}
\end{equation}
where $5^2_2$ is a non-geometric exotic brane which are obtained by two
transverse T-dualities of the NS5-brane and have been much studied in
the recent literature \cite{deBoer:2010ud, deBoer:2012ma,
Bergshoeff:2011se, Kikuchi:2012za, Andriot:2013xca, Geissbuhler:2013uka,
Kimura:2013fda, Hassler:2013wsa, Kimura:2013zva,
Chatzistavrakidis:2013jqa, Andriot:2014uda, Kimura:2013khz,
Kimura:2014upa, Okada:2014wma, Kimura:2014wga, Sakatani:2014hba,
Kimura:2014bea, Kimura:2015yla}.  This process exemplifies the fact that
standard branes can generally turn into exotic branes with non-geometric
spacetime.

The purpose of the present paper is to demonstrate how configurations
with codimension-2 sources, geometric and non-geometric, can be
represented in the 4D/5D solution.  To our knowledge, the 4D/5D solution
with codimension-2 sources has not been investigated before, and
represents a large unexplored area of research.
For the codimension{}-3 case, Eq.\ \eqref{Hharmonic} gives the general
multi-center solution.  More generally, however, the codimension-3
centers must polarize into supertubes, thus giving a multi-center
solution of codimension-3 and codimension-2 centers.  It is technically
challenging to explicitly construct general multi-center solutions
involving codimension-2 centers.  So, in this paper, we present some
simple but explicit solutions which must be useful for finding the
general solutions.
An obvious application of codimension-2 solutions is to generalize the
studies previously done for codimension-3 sources to include
codimension-2 sources mentioned above.  In \cite{deBoer:2010ud,
deBoer:2012ma}, it was argued that codimension-2 play an essential role
in the microscopic physics of black holes and we hope that this paper
will set a stage for research in that direction.

The plan of the rest of the paper is as follows.  In section
\ref{sec:setup}, we start by reviewing 5D supergravity and
the ``4D/5D solution'' which is supersymmetric and characterized by a
set of harmonic functions on $\bbR^3$.  We explain that, although
normally the harmonic functions are assumed to have codimension-3
source, they can more generally have codimension-2 source as well.  In
section \ref{codim2sol}, we present some example solutions with
codimension-2 source in the harmonic functions.  The examples include
supertubes with standard and exotic dipole charges and, in the latter
case, the spacetime is non-geometric.  In section \ref{sec:mixed}, we
give an example in which codimension-3 source and codimension-2 one
coexist.  We conclude in section \ref{sec:discussion} with remarks on
the fuzzball conjecture and the microstate geometry program.  The
appendices explain our convention and some detail of the computations in
the main text.

\section{Setup}

\label{sec:setup}

\subsection{The 4D/5D solution} \label{4D5Dsol}

%
%

We start from $d=5$, $\mathcal{N}=1$ ungauged supergravity coupled to
two vector multiplets. Including the graviphoton the theory contains
three vector fields $A^I$ ($I=1,2,3$) and two independent scalar fields
which can be parametrized by $X^I$ satisfying the constraint
$\frac{1}{6}C_{IJK}X^IX^JX^K=1$. Here, $C_{IJK}$ are constants that are
symmetric under permutations of $IJK$, and are given by
$C_{IJK}=|\epsilon_{IJK}|$ in our case.\footnote{However, most of our
expressions below are valid even for general $C_{IJK}$.} The bosonic
action of this theory is
\begin{equation}
	S=\frac{1}{16\pi G_5}\int \left(-R*1 +Q_{IJ}*F^I\wedge F^J +Q_{IJ}*dX^I\wedge dX^J -\frac{1}{6}C_{IJK}F^I\wedge F^J\wedge A^K\right)\,,\label{5Dsugraaction}
\end{equation}
where $*$ means the five-dimensional Hodge dual and $F^I=dA^I$. The metric
for the kinetic term is
\begin{equation}
	Q_{IJ}=\frac{1}{2}\operatorname{diag}\left((X^1)^{-2},(X^2)^{-2},(X^3)^{-2}\right)\,.
\end{equation}

The supersymmetric solutions of this theory have been completely
classified \cite{Gauntlett:2002nw, Gutowski:2004yv, Bena:2004de,
Gauntlett:2004qy} by solving  Killing spinor equations.  There are
two classes of supersymmetric solutions, depending on whether the Killing
vector constructed from the Killing spinor bilinear is null or
timelike. Here we will only consider the latter case.
For the timelike class solution, the metric and gauge fields are given
by
\begin{equation} \label{4D/5Dsol}
\begin{aligned}
 ds^2_5&=-Z^{-2/3}\left(dt+k\right)^2+Z^{1/3}ds_\mathrm{HK}^2\,,
 \qquad Z=Z_1 Z_2 Z_3\,,
 \\
 A^I&=B^I-Z_I^{-1}(dt+k)\,,
\end{aligned}
\end{equation}
where the functions $Z_I$ and the 1-forms $k,B^I$ depend only on the
coordinates of the 4D base with the hyper-K\"ahler metric
$ds_\mathrm{HK}^2$.  The scalars $X^I$ are
related to the electric potential $Z_I$ by
\begin{equation}
 X^I=Z^{1/3}Z_I^{-1}\,.
\end{equation}
It will be convenient to define the magnetic field
strength by
\begin{equation}
\Theta^I=dB^I\,.
\end{equation}

The demand of supersymmetry leads to the following BPS equations to be
satisfied by the quantities $\Theta^I$, $Z_I$, and $k$:
\begin{subequations}
\label{BPSeq}
\begin{align}
\Theta^I&=*_4\Theta^I \,,\label{BPSeqTheta}\\
d*_4dZ_I&=\frac{1}{2}C_{IJK}\Theta^J\wedge\Theta^K \,,\label{BPSeqZ}\\
(1+*_4)dk&=Z_I\Theta^I \,,\label{BPSeqk}
\end{align}
\end{subequations}
where $*_4$ is the Hodge dual with respect to the 4D metric
$ds^2_{\text{HK}}$.  If we solve these equations in the order presented,
the problem is linear; namely, at each step, we have a Poisson equation
with the source given in terms of the quantities found in the previous
step.

If we assume the presence of an additional translational Killing vector
that preserves the hyper-K\"ahler structure (namely, if the Killing
vector is tri-holomorphic), the 4D base should be a Gibbons-Hawking
space \cite{Gibbons:1987sp} and its metric must take the following form
\cite{Gibbons:1979zt}:
\begin{equation}
\label{GHmetric}
ds_\mathrm{HK}^2=V^{-1}(d\psi+A)^2+V\delta_{ij}dx^idx^j\,, \qquad i,j=1,2,3 \,.
\end{equation}
Here, the 1-form $A$ and the scalar $V$ depend only on the 
coordinates $x^i$ of the $\mathbb{R}^3$ base and satisfy
\begin{equation}
dA=*_3dV 
\,.
\label{AandV}
\end{equation}
The isometry direction $\psi$ has periodicity $4\pi$. The orientation of
the 4-dimensional base is given by
\begin{equation}
\epsilon_{\psi123}=+\sqrt{g_\mathrm{HK}}=V \,.
\end{equation}
From \eqref{AandV}, it is easy to see that $V$ is a \emph{harmonic} function on $\bbR^3$,
\begin{equation}
 \Lap V=0\,,\qquad
 \Lap=\partial_i \partial_i\,.\label{V_eq}
\end{equation}

\subsubsection*{Solving the BPS equations}

If we decompose $\Theta^I$ and $k$ according to the fiber-base
decomposition of the Gibbons-Hawking metric \eqref{GHmetric}, we can
solve all the BPS equations \eqref{BPSeq} in terms of harmonic functions
on $\bbR^3$.  For later convenience, let us recall how this goes in some
detail \cite{Gauntlett:2004qy}. 

First, by self-duality
\eqref{BPSeqTheta}, the 2-form $\Theta^I$ can be written as
\begin{equation}
 \Theta^I=(d\psi+A)\wedge \theta^I+V *_3 \theta^I\,,
\end{equation}
where $\theta^I$ is a 1-form on $\bbR^3$ and $*_3$ is the Hodge dual on
$\bbR^3$. The closure $d\Theta^I=0$ (the part multiplying $d\psi+A$)
implies $d \theta^I=0$, which means that $\theta^I=d\Lambda^I$ with a
scalar $\Lambda^I$.  If we plug this equation back into $d\Theta^I=0$,
we find
\begin{equation}
 \Lap (V\Lambda^I)=0\,.\label{K_eq}
\end{equation}
Therefore, $\Lambda^I=-V^{-1}K^I$ with $K^I$ harmonic, and
\begin{equation}
\Theta^I=-(d\psi+A)\wedge d(V^{-1}K^I)-V *_3d(V^{-1}K^I)\,.
 \label{Thetaharm}
\end{equation}

Next, plugging \eqref{Thetaharm} into \eqref{BPSeqZ}, we find that $Z_I$
satisfies the following Laplace equation:
\begin{align}
 \Lap Z_I
 =C_{IJK}V\, \partial_i(V^{-1}K^J)\, \partial_i(V^{-1}K^K)
 ={1\over 2}C_{IJK} \Lap (V^{-1}K^J K^K)\,,\label{Z_Ieq}
\end{align}
where in the last equality we used harmonicity of $V,K^I$.  This means
that
\begin{align}
 Z_I=L_I+{1\over 2}C_{IJK}V^{-1}K^J K^K\,,
 \label{Zharm}
\end{align}
where $L_I$ is harmonic.

Furthermore, if we decompose the 1-form $k$ as
\begin{align} \label{k_decompose}
 k=\mu (d\psi+A)+\omega\,,
\end{align}
where $\omega$ is a 1-form on $\bbR^3$,
we can show that the condition \eqref{BPSeqk} leads to another Laplace equation:
\begin{align}
 \Lap \mu = V^{-1}\partial_i [V Z_I \partial_i (V^{-1}K^I)]
 =\Lap \Bigl(
{1\over 2}V^{-1}K^I L_I+{1\over 6}C_{IJK}V^{-2}K^I K^J K^K \Bigr)\,.\label{mu_eq}
\end{align}
In the last equality, we used harmonicity of $V,K^I,L_I$.
Therefore, $\mu$ is given in terms of another harmonic function $M$ as
\begin{equation}
 \mu = M+{1\over 2}V^{-1}K^IL_I+{1\over 6}C_{IJK}V^{-2}K^I K^J K^K\,.
 \label{muharm}
\end{equation}

The 1-form $\omega$ is found by solving the equation
\begin{equation}
 *_3d\omega=VdM-MdV+\frac{1}{2}\left(K^IdL_I-L_IdK^I\right) \label{omega_eq}
\end{equation}
that also follows from \eqref{BPSeqk}.  By taking $d\,*_3$ of this
equation, we can derive the so-called integrability equation:
\begin{align}
\label{integrability}
0=V\Lap M-M\Lap V +\frac{1}{2}\left(K^I\Lap L_I-L_I\Lap
 K^I\right)\,.
\end{align}
This must be satisfied for the 1-form $\omega$ to exist. Although we
allow delta-function sources for the Laplace equations \eqref{V_eq},
\eqref{K_eq}, \eqref{Z_Ieq} and \eqref{mu_eq}, this equation
\eqref{integrability} must be imposed without allowing any delta
function in order for $\omega$ to exist.

Finally,  we note that the magnetic potential $B^I$ can be written as
\begin{equation}
B^I=V^{-1}K^I(d\psi+A)+\xi^I \,,
 \qquad
d\xi^I=-*_3dK^I \,.
\end{equation}

In summary, under the assumption of the additional $U(1)$ symmetry, we
can solve all the equations \eqref{BPSeq} in terms of harmonic functions
$V,K^I,L_I,M$.  We will refer to this solution as the ``4D/5D
solution.''

\subsubsection*{The 10 and 11-dimensional uplift}

The 5D solution \eqref{4D/5Dsol} can be thought of as coming from 11D
M-theory compactified on $T^6=T^2_{45}\times T^2_{67}\times T^2_{89}$,
with 
the following metric and the 3-form potential:
\begin{equation} \label{Msol}
\begin{split}
ds_{11}^2&=-Z^{-2/3}(dt+k)^2+Z^{1/3}ds_\mathrm{HK}^2+Z^{1/3}\left(Z_1^{-1}dx_{45}^2+Z_2^{-1}dx_{67}^2+Z_3^{-1}dx_{89}^2\right) \,,\\
\cA_3&=A^I J_I\,,\qquad
 J_1 \equiv dx^4\wedge dx^5\,,\quad
 J_2 \equiv dx^6\wedge dx^7\,,\quad
 J_3 \equiv dx^8\wedge dx^9 \,,
\end{split}
\end{equation}
where $dx_{45}^2\equiv (dx^4)^2+(dx^5)^2$ and so on.  The scalars
$X^I=Z^{1/3}Z_I^{-1}$ correspond to the volume of the 2-tori.  M-theory on
$T^6$ has $\cN=4$ supersymmetry (32 supercharges) in 5D, and the theory
\eqref{5Dsugraaction} gives its $\cN=1$ truncation in which only 8
supercharges are kept.

In the presence of the isometry direction $\psi$ in the 4D base as in \eqref{GHmetric}, the
above 11D configuration \eqref{Msol} can be reduced on it to a 10D type
IIA configuration using the formula \eqref{M-IIA} as follows:
\begin{equation}
\label{IIAfield}
\begin{split}
 ds_\mathrm{10,str}^2&=-{1\over \sqrt{V(Z-V\mu^2)}}(dt+\omega)^2+\sqrt{V(Z-V\mu^2)}\,dx^i dx^i\\
 &\qquad\qquad
 + \sqrt{Z-V\mu^2\over V}\left(Z_1^{-1}dx_{45}^2+Z_2^{-1}dx_{67}^2+Z_3^{-1}dx_{89}^2\right)\,,\\
 e^{2\Phi}&={(Z-V\mu^2)^{3/2}\over V^{3/2} Z}\,,\qquad
 B_2=\left(V^{-1}K^I-Z_I^{-1}\mu\right)J_I\,,\\
 C_1&=A-{V\mu\over Z-V\mu^2}(dt+\omega)\,,\\
 C_3&=
 \left[(V^{-1}K^I-Z_I^{-1}\mu) A+\xi^I-Z_I^{-1}(dt+\omega)
 \right]
 \wedge J_I \,.
\end{split}
\end{equation}
We note that the complexified K\"ahler moduli 
$\tau^1$, $\tau^2$, and $\tau^3$
for the 2-tori $T^2_{45}$,
$T^2_{67}$, and $T^2_{89}$, respectively, are
\begin{equation} \label{kahlerm}
\tau^1
 ={R_4 R_5\over l_s^2}\Bigl(B_{45}+i\sqrt{\det G_{ab}}\Bigr)
 ={R_4 R_5\over l_s^2}\left[\left({K^1\over V}-{\mu\over Z_1}\right)+i\frac{\sqrt{V(Z-V\mu^2)}}{Z_1V}\right]\,,
\end{equation}
where $a,b=4,5$, and similarly for $\tau^2,\tau^3$. We denoted the radii of
$x^i$ directions by $R_i, i=4,\cdots,9$.  If we compactify the theory to
4D, these $\tau^I$ become scalar moduli parametrizing the moduli space
$[SL(2,\bbR)/SO(2)]^3$.

\subsection{Codimension-3 sources}

\begin{figure}[htbp]
 \begin{quote}
  \begin{center}
   \begin{tabular}{c@{\hspace{3cm}}c}
   \includegraphics[height=4cm]{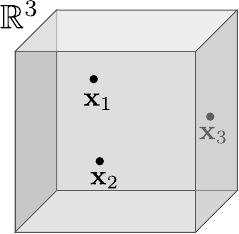} 
    &   \includegraphics[height=4cm]{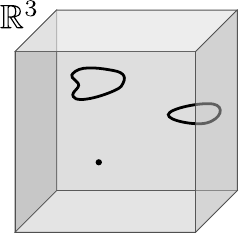}   \\
    (a) & (b) \\
   \end{tabular}
   \caption{\label{fig:sing} \sl The 4D/5D solution is specified by
   harmonic functions on the base $\bbR^3$. (a) The codimension-3
   solution is specified by point-like singularities of the harmonic
   functions.  (b) The general solution involves point-like
   (codimension-3) as well as string-like (codimension-2) singularities
   in the harmonic functions.}
  \end{center}
 \end{quote}
\end{figure}

As we have seen above, the 4D/5D solution is specified by the set of
harmonic functions $V,K^I,L_I,M$.  The general harmonic functions with
\emph{codimension-3} sources are 
\cite{Bates:2003vx, Gauntlett:2004qy}
\begin{equation}
\label{codim3harm}
\begin{aligned}
  V  &=h^0+\sum_{p=1}^N {\Gamma^0_p\over |\xv-\av_p|}\,,&\qquad
  K^I&=h^I+\sum_{p=1}^N {\Gamma^I_p\over |\xv-\av_p|}\,,\\
  L_I&=h_I+\sum_{p=1}^N {\Gamma_I^p\over |\xv-\av_p|}\,,&
  M  &=h_0+\sum_{p=1}^N {\Gamma_0^p\over |\xv-\av_p|}\,,
\end{aligned}
\end{equation}
where $\xv=(x^1,x^2,x^3)$, and $\av_p\in\bbR^3$ is the position of the
sources at which the harmonic functions become singular.  The
integrability condition \eqref{integrability} demands that the position
of the centers satisfy
\begin{align}
 \sum_{q(\neq p)}
 {\bracket{\Gamma_p,\Gamma_q} \over a_{pq}}
 =
 \bracket{h,\Gamma_p}
\end{align}
where 
$\bracket{u,v}\equiv u^0 v_0- u_0 v^0 +{1\over 2}(u^I v_I - u_I v^I)$ and
$a_{pq}\equiv |\av_p-\av_q|$.
See Fig.\ \ref{fig:sing}(a) for a schematic
explanation of codimension-3 solutions.
When we embed the 5D supergravity in
string/M-theory, these singularities are interpreted as brane sources.
For example, in the type IIA picture \eqref{IIAfield}, the singularities
in the harmonic functions \eqref{codim3harm} have the following
interpretation as brane sources \cite{Bates:2003vx}:
\begin{equation}
 V\leftrightarrow \text{D6(456789)}\,,
 \quad
 \begin{array}{l}
 K^1\leftrightarrow \text{D4(6789)}\\
 K^2\leftrightarrow \text{D4(4589)}\\
 K^3\leftrightarrow \text{D4(4567)}\\
 \end{array}\,,
 \quad
 \begin{array}{l}
 L_1\leftrightarrow \text{D2(45)}\\
 L_2\leftrightarrow \text{D2(67)}\\
 L_3\leftrightarrow \text{D2(89)}\\
 \end{array}\,,
 \quad
 M\leftrightarrow \text{D0}\,.
\label{singIIAbrn}
\end{equation}
Note that, in our description, the branes are always smeared along all
transverse directions inside the compact directions ($456789$).
For example, the D4(6789)-brane is smeared along the $45$
directions.  So, all the branes in \eqref{singIIAbrn} can be regarded as
having codimension 3 (pointlike in $\bbR^3$).

%

Many known black hole and black ring solutions in 4D and 5D are included
in the 4D/5D solutions with the harmonic functions having codimension-3
singularities, \eqref{codim3harm}.  For example, the 3-charge black hole
in 5D with the charges of M2(45), M2(67), M2(89)-branes, which is dual
to the Strominger-Vafa black hole \cite{Strominger:1996sh}, can be
expressed by the following harmonic functions:
\begin{equation}
V=\frac{1}{r}\,,\qquad K^I=0\,,\qquad L_I=1+\frac{Q_I}{r}\,,\qquad M=0\,.
\end{equation}
Other examples include the BMPV black hole \cite{Breckenridge:1996is},
the supersymmetric black ring \cite{Elvang:2004rt, Bena:2004de,
Elvang:2004ds}, the MSW black hole \cite{Maldacena:1997de}, multi-center
black hole/ring solutions \cite{Bates:2003vx} and microstate geometries
\cite{Bena:2005va, Berglund:2005vb}.

\subsection{Codimension-2 sources}

%

In the previous subsection, we considered the 4D/5D solution which has
only codimension-3 sources of D-branes.  However, recall that, in string
theory, certain combinations of branes can undergo a supertube
transition \cite{Mateos:2001qs}, under which branes spontaneously
polarize into new dipole charge, gaining size in transverse directions.
For example, as we have discussed in the Introduction, two transverse
D2-branes can polarize into an NS5-brane along an arbitrary closed curve
$\lambda$, as in \eqref{D2D2ns5intro}.  Because the NS5-brane is along a
closed curve, it has no net NS5 charge but only NS5 dipole charge. The
original D2 charges are dissolved in the NS5 worldvolume as fluxes.
When the curve $\lambda$ is inside the $\mathbb{R}^3_{123}$, which is
generically the case and is assumed henceforth, the NS5-brane appears as a
codimension-2 object in the non-compact 123 directions.
Therefore, if we are to consider \emph{generic} solutions describing D-brane
systems, we \emph{must} include codimension-2 brane sources in the
4D/5D solution.  Even in such situations,
the procedure \eqref{Thetaharm}--\eqref{muharm} to solve the BPS
equations goes through and the solution is given by the harmonic
functions $V,K^I,L_I,M$. However, they are now allowed to have
codimension-2 singularities in $\bbR^3$.
See Fig.\
\ref{fig:sing}(b) for a schematic explanation for solutions with codimension-2
sources.

To get some idea about solutions with codimension-2 sources, here we
present the harmonic functions for the $\rm D2+D2\to ns5$ supertube
\eqref{D2D2ns5intro} when the puffed-up ns5-brane is an infinite
straight line along $x^3$.\footnote{An infinitely long NS5-brane would
not be a dipole charge.  The solution \eqref{harmns5straight} must be
regarded as a near-brane approximation of an NS5-brane along a closed
curve. \label{ftnt:infinitens5}}
\begin{equation}
\label{harmns5straight} 
\begin{aligned}
 V&=1\,,\qquad
 K^1=K^2=0\,,\quad K^3=q\mspace{1mu}\theta\,,\\
 L_1&=1+Q_1\log {\Lambda\over r}\,,\quad
 L_2 =1+Q_2\log {\Lambda\over r}\,,\quad
 L_3=1\,,\qquad
 M=-{1\over 2}q\mspace{1mu}\theta\,,
\end{aligned}
\end{equation}
where $q=l_s^2/(2\pi R_8 R_9)$, $Q_1Q_2=q^2$, and $\Lambda$ is a
constant.\footnote{$\Lambda$ is the cutoff for $r$, beyond which the
near-brane approximation mentioned in footnote \ref{ftnt:infinitens5}
breaks down. } We took the cylindrical coordinates for the $\bbR^{3}$
base,
\begin{equation}
 ds_3^2=dr^2+r^2d\theta^2+(dx^3)^2\,.
\end{equation}
We will discuss such solutions more generally in the next sections.  A
novel feature is that the harmonic function $K^3$ has a branch-point
singularity along the $x^3$ axis at $r=0$.  So, $K^3$ does not just have
a codimension-2 singularity but is \emph{multi-valued}.  This $K^3$
cannot be obtained by smearing a $K^3$ with codimension-3 singularities
as in \eqref{codim3harm}.  As one can see from \eqref{IIAfield}, this
$K^3$ leads to the $B$-field
\begin{equation}
 B_2
={l_s^2 \theta\over 2\pi R_8 R_9} dx^8\wedge dx^9\,.\label{B2straightns5}
\end{equation}
Around the $x^3$-axis, this has monodromy $\Delta B_2 =l_s^2/(R_8 R_9)$,
which is the correct one for an NS5-brane extending along 34567
directions and smeared along 89 directions.  On the other hand, the
codimension-2 singularities in $L_1,L_2$ represent the D2-brane sources
dissolved in the NS5 and are obtained by smearing codimension-3
singularities in \eqref{codim3harm}.  The monodromy in $M$
\eqref{harmns5straight} does not have direct physical significance here,
because what enters in physical quantities is $\mu$, which is trivial in
the present case: $\mu=M+{1\over 2}K^3L_3=0$.

In the lower dimensional (4D) picture, the $B$-field appears as the
scalar moduli $\tau^I$ defined in \eqref{kahlerm}.
For the present case
\eqref{B2straightns5}, we have
\begin{equation}
 \tau^3={\theta\over 2\pi}\,.
\end{equation}
As we go around $r=0$, 
the modulus $\tau^3$ has the monodromy
\begin{align} \label{ns5mono}
 \tau^3\to \tau^3+1\,,
\end{align}
which can be understood as an $SL(2,\bbZ)$ duality transformation.
It was emphasized in \cite{deBoer:2010ud, deBoer:2012ma} that the charge
of the codimension-2 brane is measured by the duality monodromy around
it.  It is possible to consider codimension-2 objects around which there
is more general $SL(2,\bbZ)$ monodromy of $\tau^I$.  For example, if we
have an object around which there is the following monodromy:
\begin{align} \label{522moduli}
\tau^3 \to {\tau^3 \over -\tau^3+1}\,,
 \qquad\text{or}\quad
\tau'^3 \to \tau'^3 +1\,,
\quad \tau'^3\equiv -{1\over \tau^3}\,,
\end{align}
it corresponds to an exotic brane called the $5^2_2(34567,89)$-brane
\cite{deBoer:2010ud, deBoer:2012ma}.  This brane is non-geometric since
the $T^2_{89}$ metric is not single-valued but is twisted by a
$T$-duality transformation around it.  The $5^2_2$-brane is produced in the
supertube transition \eqref{D2D6522intro} and must also be describable
within the 4D/5D solution in terms of multi-valued harmonic functions.
We will see this in explicit examples in the following sections.


	

\section{Examples of codimension-2 solutions} \label{codim2sol}

In the previous section, we have motivated codimension-2 solutions and
presented simplest examples of them -- straight supertubes.  In this
section, we consider more ``realistic'' codimension-2 solutions that
should serve as building blocks for constructing more general solutions.

\subsection{1-dipole solutions}


We begin with the case of a pair of D-branes puffing up into a supertube
with one new dipole charge, such as \eqref{D2D2ns5intro} and
\eqref{D2D6522intro} presented in the Introduction.  The supergravity
solution for such 1-dipole supertubes can be obtained by dualizing the
known solutions describing supertubes, such as the one in
\cite{Emparan:2001ux}.\footnote{See e.g.\ \cite{Lunin:2001fv,
deBoer:2012ma} for details of such dualization
procedures.\label{ftnt:dualization}} In that sense, the solutions
presented here are not new.  However, they have not been discussed in
the context of the 4D/5D solutions and harmonic functions as we do here.

\subsubsection*{D2(67)+D2(45)$\to$ns5($\lambda$4567)}
	
As just mentioned, the supergravity solution for the $\rm D2+D2\to
ns5$ supertube \eqref{D2D2ns5intro} can be obtained by dualizing known
solutions, and we can read off from it the harmonic functions using the
relations in the previous section.
Explicitly, the harmonic functions are
\begin{equation} \label{harmonicD2D2}
\begin{aligned}
V&=1\,,\quad&
K^1&=0\,,& K^2&=0\,,& K^3&=\gamma\,,\\
&&L_1&=f_2\,,& L_2&=f_1\,,& L_3&=1\,,&
\quad M&=-\frac{\gamma}{2}\,.
\end{aligned}
\end{equation}
Here, the harmonic functions $f_1$ and $f_2$ are given by
\begin{gather}
 f_1=1+\frac{Q_1}{L}\int_0^L \frac{d\lambda}{|\xv-\Fv(\lambda)|}\,,\qquad f_2=1+\frac{Q_1}{L}\int_0^L \frac{|\dot\Fv(\lambda)|^2d\lambda}{|\xv-\Fv(\lambda)|}\,,\label{f1f2_for_D2D2ns5}
\end{gather}
where $\mathbf{x}=\mathbf{F}(\lambda)$ is the profile of the supertube in
$\mathbb{R}^3$ and satisfies
$\mathbf{F}(\lambda+L)=\mathbf{F}(\lambda)$.  The functions $f_1$ and
$f_2$ represent the D2(67) and D2(45) charges, respectively, dissolved
in the codimension-2 worldvolume of the ns5 supertube.  $Q_1$ is the
D2(67) charge, while the D2(45) charge is given by
\begin{gather}
 Q_2=\frac{Q_1}{L}\int_0^L d\lambda\,\abs{\dot{\mathbf{F}}(\lambda)}^2\,.
\end{gather}
The charges $Q_1,Q_2$ are related to the quantized D-brane numbers $N_1,N_2$ by
\begin{equation} \label{charges_d2d2ns5}
Q_1=\frac{g_sl_s^5}{2R_4R_5R_8R_9}N_1\,,\qquad Q_2=\frac{g_sl_s^5}{2R_6R_7R_8R_9}N_2\,,\qquad
L=\frac{2\pi g_sl_s^3}{R_4R_5}N_1\,.
\end{equation}
where $R_i$, $i=4,\dots,9$ are the radii of the $x^i$ directions. We
have also written down the expression for $L$, the periodicity of the
profile function $\Fv(\lambda)$, in terms of other
quantities.\footnote{In the F1-P system, $L$ corresponds to the length
of the fundamental string.  For the expressions of $L$ in different
duality frames, see references in footnote \ref{ftnt:dualization}.  }

\begin{figure}[htbp]
 \begin{quote}
  \begin{center}
   \vspace*{.2cm}
   \includegraphics[height=3cm]{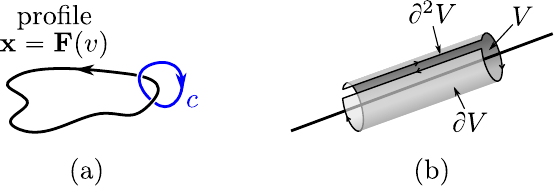} \caption{\label{fig:cycle}
   \sl (a) The function $\gamma$ has a monodromy as one goes around the
   cycle $c$ that links with the profile.  (b) The integral region in
   Eq.\ \eqref{int_Lap_gamma}.  The contribution from the top and bottom
   surfaces of the tube is negligible if the tube is very thin.}
  \end{center}
 \end{quote}
\end{figure}

The function $\gamma$ is defined via the differential equation
\begin{equation}
d\alpha=*_3d\gamma\label{alpha_gamma}
\end{equation}
where $\alpha$ is a 1-form in $\mathbb{R}^3$ given by (see Appendix \ref{app:alpha})
\begin{align}
\alpha_i=\frac{Q_1}{L}\int_0^L \frac{\dot F_i(\lambda)\,d\lambda}{|\xv-\Fv(\lambda)|}\,.\label{alpha_def}
\end{align}
It is easy to see from \eqref{alpha_gamma} that
$\gamma$ is  harmonic: $\Lap
\gamma=*_3d*_3d \gamma = *_3 d^2 \alpha= 0$.
Note that, even though $\alpha$ is single-valued, the function $\gamma$
defined via the differential equation \eqref{alpha_gamma} is
multi-valued and has a monodromy as we go along a closed circle $c$ that
links with the profile; see Fig.\ \ref{fig:cycle}(a).  The monodromy of
$\gamma$ can be computed by integrating $d\gamma$ along $c$, which can
be homotopically deformed to a very small circle near some point on the
profile, and is equal to
\begin{equation} \label{dgamma}
\int_c d\gamma
 =\int_c *_3d\alpha
 =\frac{4\pi Q_1}{L}\,.
\end{equation}

The integrability condition \eqref{integrability} requires
\begin{align}
V\Lap M-M\Lap V
 +\frac{1}{2}\left(K^I\Lap L_I-L_I\Lap K^I\right)
 =-\Lap \gamma\equiv 0\,.\label{int_cond_1dip}
\end{align}
Superficially, this is satisfied because $\gamma$ is harmonic.  However,
one must be careful because $\gamma$ is singular along the profile and
may have delta-function source there (as is the case for $L_{1,2}$).  We
can show that it actually does not even have delta-function source as
follows.  If we integrate $\Lap \gamma$ over a small tubular volume $V$
containing the profile $\xv=\Fv(\lambda)$, we get
\begin{align}
 \int_{V} d^3x\,\, \Lap \gamma
 =\int_{V} d {*_3 d \gamma}
 =\int_{\partial V} *_3 d \gamma
 =\int_{\partial V} d \alpha
 =\int_{\partial^2 V} \alpha
 =0\,,\label{int_Lap_gamma}
\end{align}
where the last equality holds because $\alpha$ is single-valued.  See
Fig.\ \ref{fig:cycle}(b) for explanation of the integral region.
Therefore, $\Lap \gamma$ in \eqref{int_cond_1dip} vanishes everywhere,
even on the profile, and the integrability condition is satisfied for
any profile $\Fv(\lambda)$.

From harmonic functions \eqref{harmonicD2D2}, we can read off various
functions and forms that appear in the full  solution:
\begin{equation}
 Z_1=f_2\,,\quad Z_2=f_1\,,\quad Z_3=1\,,
 \quad\mu=0\,,\quad \omega=-\alpha\,,\quad
 \xi^1=\xi^2=0\,,\quad\xi^3=-\alpha\,.
\end{equation}
The existence of $\omega$ is guaranteed by the integrability condition.
Substituting this data into \eqref{IIAfield}, we obtain the type IIA
fields:
\begin{align}
 \label{D2D5ns5IIA}
\begin{split}
 ds_{10}^2&=-(f_1f_2)^{-1/2}(dt-\alpha)^2+(f_1f_2)^{1/2}dx^idx^i \\
	&\qquad +(f_1/f_2)^{1/2}dx_{45}^2+(f_2/f_1)^{1/2}dx_{67}^2+(f_1f_2)^{1/2}dx_{89}^2 \,,\\
 e^{2\Phi}&=(f_1f_2)^{1/2}\,,\qquad B_2=\gamma\,dx^8\wedge dx^9\,,\\
 C_1&=0\,,\qquad
 C_3=-f_2^{-1}(dt-\alpha)\wedge dx^4\wedge dx^5-f_1^{-1}(dt-\alpha)\wedge dx^6\wedge dx^7\,,\\
\end{split}
\end{align}
where we have dropped some total derivative terms in the RR potentials.
Since $f_1,f_2\to 1$ as $|\xv|\to \infty$, the spacetime is
asymptotically $\bbR^{1,3}\times T^6$.  Multi-valuedness is restricted
to the $B$-field and the metric is single-valued; namely, this solution
is geometric.


One can show that the solution \eqref{D2D5ns5IIA} has the expected
monopole charge; it has monopole charge for D2(67) and D2(45) but not
for NS5 (we show this for more general solutions in the next
subsection).  The dipole charge for NS5 is easier to see in the
monodromy of the K\"ahler moduli, as we discussed around
\eqref{B2straightns5}, and their values are
\begin{equation}
\tau^1=i\frac{R_4R_5}{l_s^2}\sqrt{f_1\over f_2}\,,\qquad \tau^2=i\frac{R_6R_7}{l_s^2}\sqrt{f_2\over f_1}\,,\qquad \tau^3= \frac{R_8R_9}{l_s^2}\left(\gamma+i\sqrt{f_1f_2}\right)\,.
\end{equation}
$\tau^1$ and $\tau^2$ are single-valued while, as we can see from
\eqref{dgamma}, $\tau^3$ has the following monodromy
as we go around the supertube along cycle $c$:
\begin{equation}
\tau^3\to\tau^3+1\,,
\end{equation}
where we used \eqref{charges_d2d2ns5} and \eqref{dgamma}.
This is the correct monodromy around an NS5-brane.  So, this
solution has the expected monopole and dipole charge.

Although we have derived the harmonic functions \eqref{harmonicD2D2} by
dualizing known solutions, we can also derive it by requiring that they
represent the charge and dipole charge expected of the supertube
\eqref{D2D2ns5intro} as follows.  First, no D6-brane means $V=1$ and no
D0-brane means $\mu=0$.  Then \eqref{IIAfield} implies that, in order to
have an NS5-brane along the profile $\Fv(\lambda)$, the harmonic
function $K^3\equiv \gamma$ must have the monodromy \eqref{dgamma}.  As
we show in Appendix \ref{app:alpha}, this means that $\gamma$ must be
given in terms of $\alpha$ via \eqref{alpha_gamma} and \eqref{alpha_def}.
Next, to account for the D2 charges dissolved in the NS5 worldvolume, we
need $L_1,L_2$ given in \eqref{harmonicD2D2} and
\eqref{f1f2_for_D2D2ns5}.

	
Note that, if we lift the supertube \eqref{D2D2ns5intro}
to M-theory, we have
\begin{equation}
\mathrm{M2(67)+M2(45)\to m5(\lambda4567)}\,.
\end{equation}
Therefore, our solution simply corresponds to the 4D version of Bena and
Warner's solution in \cite{Bena:2004de}. The difference is that they
were discussing 5D solutions with general supertube shapes, while we are
focusing on solutions which has an isometry and can be reduced to 4D\@.
Because of that, we can be more explicit in the solution in terms of
harmonic functions.



\subsubsection*{D2(89)+D6(456789)$\to5_2^2(\lambda$4567;89)}

The second example is the $\rm D2+D6\to 5^2_2$ supertube
\eqref{D2D6522intro}, which can be obtained by taking the $T$-dual of
the above solution \eqref{D2D5ns5IIA} along 6789 directions.  Involving
the exotic $5^2_2$-brane, this is a non-geometric supertube where the
metric becomes multi-valued.\footnote{The metric for an exotic
non-geometric supertube ($\rm D4+D4\to 5^2_2$) was first discussed in
\cite{deBoer:2010ud, deBoer:2012ma}.}


Harmonic functions which describe this supertube 
\eqref{D2D6522intro} are
\begin{equation}
\begin{aligned}
V&=f_2\,,\quad 
 &K^1&=\gamma\,,& K^2&=\gamma\,,& K^3&=0\,,\\
&&L_1&=1\,,     & L_2&=1\,,     & L_3&=f_1\,,&\quad
M&=0\,.
\end{aligned}
\end{equation}
The charges appearing in harmonic functions are related to 
brane numbers by
\begin{equation}
Q_1=\frac{g_sl_s^5}{2R_4R_5R_6R_7}N_1\,,\qquad Q_2=\frac{g_sl_s}{2}N_2\,,\qquad L=\frac{2\pi g_sl_s^7}{R_4R_5R_6R_7R_8R_9}N_1\,.
\label{charges_d2d6522}
\end{equation}
As we can easily check, the integrability condition
\eqref{integrability} is trivially satisfied.
The various functions and forms are
\begin{equation}
Z_1=Z_2=1\,,\quad Z_3=f_1F\,,\quad \xi^1=\xi^2=-\alpha\,,\quad \xi^3=0\,,\quad \mu=f_2^{-1}\gamma\,,\quad\omega=-\alpha\,.
\end{equation}
The IIA fields are given by
\begin{align}
 \label{D2D6522IIA}
\begin{split}
 ds_{10}^2&=-(f_1f_2)^{-1/2}(dt-\alpha)^2+(f_1f_2)^{1/2}dx^idx^i+(f_1/f_2)^{1/2}\left(dx_{4567}^2+f_1^{-1}F^{-1}dx_{89}^2\right)\,,\hspace*{-1cm}\\
	e^{2\Phi}&=f_1^{1/2}f_2^{-3/2}F^{-1}\,,\qquad 	B_2=-\frac{\gamma}{f_1f_2F}\,dx^8\wedge dx^9,\\
 C_1&=\beta_2-f_1^{-1}\gamma\,(dt-\alpha)\,,\\
 C_3&=-\frac{1}{f_1F}(dt-\alpha)\wedge dx^8\wedge dx^9-\frac{\gamma}{f_1f_2F}\,\beta_2\wedge dx^8\wedge dx^9\,,
\end{split}
\end{align}
where we defined
\begin{equation}
F \equiv 1+\frac{\gamma^2}{f_1f_2}\,.
\end{equation}
We have dropped some total derivative terms in the RR potentials.  Since
$f_1,f_2\to 1$ as $|\xv|\to \infty$, the spacetime is asymptotically
$\bbR^{1,3}\times T^6$.  However, because the multi-valued function
$\gamma$ enters the metric, this spacetime is non-geometric.  Every time
one goes through the supertube, one goes to different spacetime with
different radii for $T^2_{89}$, although it is related to the original
one by $T$-duality.

It is not difficult to show that the solution \eqref{D2D6522IIA} carries
the expected monopole charge for D2(89) and D6(456789), and not for
other charges.  To see the $5^2_2$ dipole charge, let us look at the
K\"ahler moduli which are
\begin{equation}
\tau^1=i\frac{R_4R_5}{l_s^2}\sqrt{\frac{f_1}{f_2}}\,,\quad
 \tau^2=i\frac{R_6R_7}{l_s^2}\sqrt{\frac{f_1}{f_2}}\,,\quad 
\tau^3=\frac{R_8R_9}{l_s^2}\left(-\frac{\gamma}{f_1f_2F}+i\frac{1}{\sqrt{f_1f_2}F}\right)\,.
\end{equation}
If we define
\begin{equation}
\tau'^3\equiv -\frac{1}{\tau^3}=\frac{l_s^2}{R_8R_9}\left(\gamma+i\sqrt{f_1f_2}\right) \,,
\end{equation}
the monodromy  around the supertube is simply
\begin{equation}
\tau'^3\to\tau'^3+1\,,
\end{equation}
where we used \eqref{dgamma} and \eqref{charges_d2d6522}.  This is the
correct monodromy for the $5^2_2$-brane.

Although one sees that the RR potentials are also multi-valued in
\eqref{D2D6522IIA}, this does not mean that we have further monopole or
dipole charges.   We will see this in a different example in
subsection \ref{ss:2-dip}.

\subsubsection*{Other duality frames}

One can also consider supertube transitions in other duality frames,
such as
\begin{align}
 \rm D0+D4(4567)\to ns5(\lambda 4567)
 \label{D0D4_ns5}
\end{align}
or
\begin{align}
 \rm D4(6789)+D4(4589)\to 5^2_2(\lambda 4567,89)\,.
 \label{D4D4_522}
\end{align}
The latter transition \eqref{D4D4_522} was studied in
\cite{deBoer:2010ud, deBoer:2012ma}.  The configuration on the left hand
side of \eqref{D0D4_ns5} and \eqref{D4D4_522} are not in the timelike
class but in the null class \cite{Gauntlett:2002nw, Gutowski:2004yv},
and their analysis requires a different 5D ansatz from the one we used
above.


\subsection{2-dipole solutions}
\label{ss:2-dip}



\medskip
\subsubsection*{A naive attempt}

In the above, we demonstrated how the codimension-2 solution with one
dipole charge fits into the 4D/5D solution. The next step is to combine
two such solutions so that there are two different types of dipole
charge.
For example, can we construct a solution in which the supertube
transition \eqref{D2D2ns5intro} happens simultaneously for two different
D2-D2 pairs?  For example, consider
\begin{equation} \label{D2D2D2ns5ns5}
\begin{array}{lclcl}
\mathrm{D2(45)}&+&\mathrm{D2(89)}&\to&\mathrm{ns5(\lambda 4589)}\\
\mathrm{D2(67)}&+&\mathrm{D2(89)}&\to&\mathrm{ns5(\lambda 6789)}
\end{array}\,.
\end{equation}
How can we construct harmonic functions corresponding to this
configuration?  For co\-di\-men\-sion-3 solutions \eqref{codim3harm},
having multiple centers was achieved just by summing the harmonic
functions for each individual center. So, a naive guess is to simply sum
the harmonic functions for each individual supertube, as
follows:\footnote{This was obtained by permuting $K^I,L_I$ of
\eqref{harmonicD2D2} and also by a suitable reparametrization of
$\lambda$ in $f_1',f_2'$.}
\begin{equation}
\begin{aligned}
V&=1\,,\quad&
K^1&=\gamma'\,,& K^2&=\gamma\,,& K^3&=0\,,\\
&&L_1&=f_1\,,& L_2&=f_1'\,,& L_3&=f_2+f_2'\,,&\quad M&=-\frac{\gamma}{2}-\frac{\gamma'}{2}\,.
\end{aligned}\label{2tubeharmonic_naive}
\end{equation}
However, this does not work; as one can easily check, the integrability
condition \eqref{integrability} is not generally satisfied for this
ansatz \eqref{2tubeharmonic_naive}. The two dipoles talk to each other
and we must appropriately modify the harmonic functions to construct a
genuine solution.

\subsubsection*{A non-trivial 2-dipole solution}

So, the above naive attempt does not work and we must take a different
route to find a 2-dipole solution.  Here, we use the superthread (or
supersheet) solution of \cite{Niehoff:2012wu} to construct one.  The
superthread solution describes a system of D1 and D5-branes with
traveling waves on them, and corresponds to the following simultaneous
supertube transitions:
\begin{equation} \label{D1D5P}
\begin{array}{lclcl}
\mathrm{D1(5)}&+&\mathrm{P(5)}&\to&\mathrm{d1(\lambda)}\\
\mathrm{D5(56789)}&+&\mathrm{P(5)}&\to&\mathrm{d5(\lambda 6789)}
\end{array}\,.
\end{equation}
The left hand side of \eqref{D1D5P} can be thought of as the
constituents of the 3-charge black hole. This is  not just a trivial
superposition of D1-P and D5-P supertubes, since the two supertubes
interact with each other.

The superthread solution was originally obtained as a BPS solution in 6D
supergravity. The BPS equations in 6D have a linear structure
\cite{Bena:2011dd} which descends to that of the 5D equations
\eqref{BPSeq} and facilitates the construction of explicit solutions.
The 6D BPS equations involve a lightlike coordinate $v$ and a
4-dimensional base space which is flat $\mathbb{R}^4$ for the
superthreads.  We use $\vec{x}=(x^1,x^2,x^3,x^4)$ for the coordinates of
$\bbR^4$.
The superthread solution is characterized by profile functions
$\vec{F}_p(v)$, which describe the fluctuation of the D1 and D5-brane
worldvolume. The index $p=1,\cdots,n$ labels different threads of the
D1-D5 supertubes. We review the superthread solution in Appendix
\ref{app:superthread}.

If we smear the superthread solution along $x^4$ and $v$ directions, it
describes the D1-D5-P supertube \eqref{D1D5P} extending along the
$\bbR^3_{123}$ directions and can be connected to the 4D/5D solutions
discussed in section \ref{4D5Dsol}.  After duality
transformations,\footnote{Specifically, to go from \eqref{D1D5P} to
\eqref{D2D2D2ns5ns5}, we can take $T_{4567}$, $S$, then $T_4$ duality
transformations and rename coordinates as $456789\to 894567$, so that
D1(5), D5(56789), P(5) charges map into D2(45), D2(67), D2(89) charges,
respectively.  } the resulting solution can be regarded as describing
precisely the 2-dipole configuration \eqref{D2D2D2ns5ns5}.
More precisely, the final configuration is as follows.  We have $n$
supertubes labeled by $p=1,\dots,n$ and the $p$-th tube has the profile
$\xv=\Fv_p(\lambda_p)\in\bbR^3$, where $\lambda_p$ parametrizes the
profile and the function $\Fv_p$ has the periodicity
$\Fv_p(\lambda_p+L_p)=\Fv_p(\lambda_p)$.  The $p$-th tube carries the
D2(45), D2(67), D2(89) monopole charges $Q_{p1}, Q_{p2}, Q_{p3}$
respectively, as well as ns5 dipole charges displayed in \eqref{D2D2D2ns5ns5}.

Explicitly, the harmonic functions describing the 2-dipole configuration
\eqref{D2D2D2ns5ns5} are
\begin{subequations}
\label{2tubeharmonic}
\begin{align}
V&=1 \,,\qquad K^1=\gamma_2\,,\quad K^2=\gamma_1\,,\quad K^3=0\,,
\label{2tubeharmonic_VK}\\[1ex]
L_I&=1+\sum_p{Q_{pI}}\int_p\frac{1}{R_p}=Z_I\,,\qquad I=1,2\,,\label{2tubeharmonic_L1L2}\\[1ex]
L_3&=1+\sum_p\int_p\frac{\rho_p}{R_p}\notag\\[-1ex]
 &\qquad
 +\sum_{p,q}{Q_{pq}}\iint_{p,q}\left[\frac{\dot{\mathbf{F}}_p\cdot\dot{\mathbf{F}}_q}{2R_pR_q}-\frac{\dot{F}_{pi}\dot{F}_{qj}(R_{pi}R_{qj}-R_{pj}R_{qi})}{F_{pq}R_pR_q(F_{pq}+R_p+R_q)}\right]-K^1 K^2\,,\hspace*{-1cm}
\label{2tubeharmonic_L3}
 \\[1ex]
M&={1\over 2}\sum_{p,q}{Q_{pq}}\iint_{p,q}\frac{\epsilon_{ijk}\dot{F}_{pqi}R_{pj}R_{qk}}{F_{pq}R_pR_q(F_{pq}+R_p+R_q)}
 -\frac{1}{2}(K^1L_1+K^2L_2)\label{2tubeharmonic_M}
\end{align}
\end{subequations}
where we defined
\begin{align}
\begin{split}
 \Rv_p(\lambda_p)&\equiv \mathbf{x}-\mathbf{F}_p(\lambda_p)\,,\qquad
 \Fv_{pq}(\lambda_p,\lambda_q)\equiv \mathbf{F}_p(\lambda_p)-\mathbf{F}_q(\lambda_q)\,,\\
 R_p&\equiv \abs{\Rv_p}\,,\qquad 
 F_{pq}\equiv \abs{\Fv_{pq}}\,,\qquad\qquad
 Q_{pq}\equiv Q_{p1}Q_{q2}+Q_{p2}Q_{q1}\,.
\end{split}\label{def_R,F,Qpq}
\end{align}
Also, for integrals along the supertubes, we defined
\begin{equation}
 \int_p \equiv \frac{1}{L_p}\int_0^{L_p} d\lambda_p\,,\qquad
 \iint_{p,q} \equiv \frac{1}{L_p L_q}\int_0^{L_p} d\lambda_p
 \int_0^{L_q} d\lambda_q\label{def_int_p}
\end{equation}
and the dependence on the parameter $\lambda_p$ in \eqref{2tubeharmonic}
has been suppressed.\footnote{For example, the first term in the second
line of \eqref{2tubeharmonic_L3} means
\begin{math}
 \sum_{p,q=1}^n{Q_{pq}\over L_p L_q}\int_0^{L_p} d\lambda_p \int_0^{L_q} d\lambda_q
 \frac{\dot{\mathbf{F}}_p(\lambda_p)\cdot\dot{\mathbf{F}}_q(\lambda_q)}{2R_p(\lambda_p)R_q(\lambda_q)}.
\end{math}
Note that, even for $p=q$, the integral is two-dimensional; namely, the
summand for $p=q$ is
\begin{math}
 {Q_{pp}\over L_p^2}\int_0^{L_p} d\lambda_p \int_0^{L_p} d\lambda_p'
 \frac{\dot{\mathbf{F}}_p(\lambda_p)\cdot\dot{\mathbf{F}}_p(\lambda_p')}{2R_p(\lambda_p)R_p(\lambda_p')}.
\end{math}
} The quantity $\rho_p(\lambda_p)$ in \eqref{2tubeharmonic_L3} is an
arbitrary function corresponding to the D2(89) density along the $p$-th
tube.  A similar density could be introduced for $M$ in
\eqref{2tubeharmonic_M}, but it had been ruled out by a no-CTC (closed
timelike curve) analysis in \cite{Niehoff:2012wu} and was not included
here.
The scalars $\gamma_I$ satisfy
\begin{equation}
 d\gamma_I=*_3 d\alpha_I\,,\qquad
 \alpha_I=\sum_p Q_{pI}\int_p 
    \frac{\dot{\Fv}_{p} \cdot d\xv}{R_p}\,,\qquad I=1,2\,,
\end{equation}
generalizing \eqref{alpha_gamma}, \eqref{alpha_def}.  
Furthermore, the 1-form $\omega$ is given by
\begin{subequations}
\begin{align}
 \omega&=\omega_0+\omega_1+\omega_2\,,\\
 \omega_0&=
 \sum_{p}
 (Q_{p1}+Q_{p2})\int_p {\dot{\Fv}_{p} \cdot d\xv \over  R_p}\,,
 \qquad
 \omega_1=
 {1\over 2}\sum_{p,q}Q_{pq}
 \iint_{p,q} {\dot{\Fv}_{p}\cdot d\xv \over R_p R_q}\,,
 \\
 \omega_2&=
 {1\over 4}\sum_{p,q}Q_{pq}\iint_{p,q}
 {\dot{F}_{pqi}\over F_{pq}}
 \Biggl[
 \left({1\over R_p}-{1\over R_q} \right)dx^i
 -
 2{R_{pi}R_{qj}-R_{pj}R_{qi}
 \over R_p R_q(F_{pq}+R_p+R_q)}
 dx^j
 \Biggr].
\end{align}
\label{omega_2-tube}
\end{subequations}
The charges
$Q_{pI},Q_{p3}$ and the profile length $L_p$ are related to quantized
numbers by\footnote{The $p$-th tube has equal D2(45) and D2(67) numbers
by construction.  It is also possible for the $p$-th tube to carry only
the D2(45) (or D2(67)) charge.  In that case, $Q_{p2}=0$ (resp.\
$Q_{p1}=0$) and $Q_{p1}$ ($Q_{p2}$) is still given by
\eqref{charges_d2d2d2ns5ns5}.}
\begin{align} 
\begin{split}
 Q_{p1}&=\frac{g_s l_s^5}{2R_6 R_7 R_8 R_9}N_p\,,\qquad 
 Q_{p2} =\frac{g_s l_s^5}{2R_4 R_5 R_8 R_9}N_p\,,\\
 Q_{p3}&=\frac{g_s l_s^5}{2R_4 R_5 R_6 R_7}N_{p3}\,,\qquad
 L_p=\frac{2\pi g_sl_s^3}{R_4R_5}N_p\,.
\end{split}
 \label{charges_d2d2d2ns5ns5}
\end{align}

It is interesting to compare the above harmonic functions
\eqref{2tubeharmonic} with the naive guess \eqref{2tubeharmonic_naive}.
The naive $V,K^1,K^2,K^3,L_1,L_2$ were correct, but $L_3,M$ needed
correction terms proportional to $Q_{pq}$ to be a genuine solution.
Since $Q_{pq}$ involves the product of two types of charge (D2(45) and
D2(67)) and represents interaction between two different dipoles.

It is not immediately obvious that $L_3$ and $M$ in
\eqref{2tubeharmonic} are harmonic on $\mathbb{R}^3$.  One can show that
their Laplacian is given by
\begin{align}
 \Lap L_3&=-4\pi\sum_p \int_p \rho_p\,\delta^3(\xv-\Fv_p)
 -4\pi \sum_{p,q}Q_{pq}\iint_{p,q} {\dot{\Fv}_p\cdot \dot{\Fv}_q\over F_{pq}}
 \,\delta^3(\xv-\Fv_p)
 \,,\label{LapL3}\\
 \Lap M&
 =-{1\over 2}K^I\Lap L_I
 =2\pi \sum_p Q_{pI}\int_p K^I(\Fv_p)\,\delta^3 (\xv-\Fv_p)\,.
 \label{LapM}
\end{align}
Namely, $L_3$ and $M$ are harmonic up to delta-function source along the
profile.  In deriving these, we used the following relations:
\begin{gather}
\Lap\left[\frac{R_{pi}R_{qj}-R_{pj}R_{qi}}{F_{pq}R_pR_q(F_{pq}+R_p+R_q)}\right]
=
-\frac{R_{pi}R_{qj}-R_{pj}R_{qi}}{R_p^3R_q^3}\,,\\
\int_p {\Rv_{p}\cdot\dot{\Fv}_{p}\over R_p^3}
=\int_p \partial_{\lambda_p}\!\!\left({1\over R_p}\right)=0\,,
 \qquad
 \Lap\! \left({1\over |\xv|}\right)=-4\pi\delta^3(\xv)\,.
\end{gather}
With the relations \eqref{LapL3} and \eqref{LapM}, it is straightforward
to show that the integrability condition \eqref{integrability} is
identically satisfied for any profile.

The harmonic functions $L_3,M$ in \eqref{2tubeharmonic} are
multi-valued, because $K^1,K^2$ are.  However, the quantities that
actually enter the 10D metric \eqref{IIAfield} are single-valued.
Indeed,
\begin{subequations}
\label{Z3_mu_D2D2D2ns5ns5}
\begin{align}
Z_3&=1+\sum_p\int_p\frac{\rho_p}{R_p}
 +\sum_{p,q}{Q_{pq}}\iint_{p,q}\left[\frac{\dot{\mathbf{F}}_p\cdot\dot{\mathbf{F}}_q}{2R_pR_q}-\frac{\dot{F}_{pi}\dot{F}_{qj}(R_{pi}R_{qj}-R_{pj}R_{qi})}{F_{pq}R_pR_q(F_{pq}+R_p+R_q)}\right]\,,
\label{Z3_D2D2D2ns5ns5}
 \\
\mu&={1\over 2}\sum_{p,q}{Q_{pq}}\iint_{p,q}\frac{\epsilon_{ijk}\dot{F}_{pqi}R_{pj}R_{qk}}{F_{pq}R_pR_q(F_{pq}+R_p+R_q)}\,.\label{mu_D2D2D2ns5ns5}
\end{align}
\end{subequations}
So, the metric is single-valued and the spacetime is geometric.  This is
as it should be because the configuration \eqref{D2D2D2ns5ns5} does not
contain any non-geometric exotic branes.

\subsubsection*{Single/multi-valuedness and physical condition}

It is instructive to see how these multi-valued harmonic functions come
about in solving the BPS equations as reviewed in subsection
\ref{4D5Dsol}.  Assume that we are given $V,K^I$ of
\eqref{2tubeharmonic_VK} (which corresponds to having specific ns5-brane
dipole charges and no D6-brane), and consider finding $L_I,M$ or
equivalently $Z_I,\mu$ from the BPS equations.  To find $Z_I$, we must
solve \eqref{Z_Ieq}.  For $I=1,2$, this gives a simple Laplace equation
for $L_1,L_2$, whose solution is \eqref{2tubeharmonic_L1L2}.  On
the other hand, the equation \eqref{Z_Ieq} for $Z_3$ reads
\begin{align}
 \Lap Z_3=\Lap(K^1 K^2)=2\partial_i K^1 \partial_i K^2
 =2(\partial_i \alpha_{1j} \partial_i \alpha_{2j}
   -\partial_i \alpha_{1j} \partial_j \alpha_{2i})\,.\label{Lap_Z3}
\end{align}
Although $K^{1,2}$ are multi-valued, the last expression in
\eqref{Lap_Z3} is a single-valued.  Therefore, it is possible to solve
this Poisson equation for $Z_3$ using the standard Green function
$-{1\over 4\pi}{1\over |\xv-\xv'|}$, and the result will be
automatically single-valued.  The above solution \eqref{Z3_D2D2D2ns5ns5}
corresponds to this solution. This is physically the correct solution in
the current situation where we only have standard (D2 and NS5) branes
and the metric must be single-valued.
Alternatively, we can solve \eqref{Lap_Z3} in terms of a multi-valued
function.  If we rewrite \eqref{Lap_Z3} as $\Lap L_3=0$ with
$L_3=Z_3-K^1 K^2$, then $L_3=1+\sum_p\int_p (\rho_p/R_p)\equiv
L_3^{\text{alt}}$ is a possible solution.  This is the direct analogue
of what we did for the codimension-3 solution.  This gives a
multi-valued $Z_3=L_3+K^1 K^2\equiv Z_3^{\text{alt}}$ and hence a
multi-valued metric, which is physically unacceptable.

One may find it strange that there are two different solutions, $Z_3$ of
\eqref{Z3_D2D2D2ns5ns5} and $Z_3^{\text{alt}}$, to the same Poisson
equation \eqref{Lap_Z3}.  However, the solution to the Poisson equation
is unique given the boundary condition at infinity.  The two solutions
have different boundary conditions (a single-valued one for the $Z_3$ of
\eqref{Z3_D2D2D2ns5ns5} and a multi-valued one for $Z_3^{\text{alt}}$)
and there is no contradiction that they are both solutions to the same
Poisson equation.  The BPS equations such as \eqref{Lap_Z3} must be
solved taking into account the physical situation one is considering.

The $\mu$ equation \eqref{mu_eq} is
\begin{align}
 \Lap \mu 
= {1\over 2}\Lap(K^I L_I)
= \partial_i K^I\partial_i Z_I 
= \epsilon_{ijk}|\epsilon_{IJ}| \partial_j \alpha_{Jk}\partial_i Z_I \,.
\end{align}
Again, we have two options.  The first one is to use the standard
single-valued Green function to the last expression to obtain the
single-valued $\mu$ as given in \eqref{mu_D2D2D2ns5ns5}.  The second one
is to rewrite the above as $\Lap M=0$, $M=\mu - (1/2)K^IL_I$ and say
that $M$ is single-valued.  This gives multi-valued $\mu$ and is
inappropriate for the current situation.

\subsubsection*{Closed timelike curves}

It is known that near an over-rotating supertube there can be closed
timelike curves (CTCs) which must be avoided in physically acceptable
solutions \cite{Emparan:2001ux, Niehoff:2012wu}.  The dangerous
direction for the CTCs is known to be along the supertube, which is
inside $\bbR^3$.
%
%
By setting $dt=d\psi=0$ in the metric \eqref{4D/5Dsol}, the line element
inside $\bbR^3$ is
\begin{align}
dl^2=-Z^{-2/3}(\mu A+\omega)^2+Z^{1/3}(V^{-1}A^2+Vd\xv^2)\,.
\end{align}
In the present case, we have $V=1$ and $A=0$, and therefore the line
element becomes
\begin{equation} \label{sgi16Oct14}
dl^2= Z^{-2/3}(-\omega^2+Zd\xv^2)\,,
\end{equation}
where $\omega$ is given by 
\eqref{omega_2-tube}.

In the near-tube limit in which we approach a particular point
$\Fv_p(\lambda_p^0)$ on the $p$-th curve, where $\lambda_p^0$ is the
value of the parameter corresponding to that point, the functions
$Z_{1,2,3}$ can be expanded as
\begin{subequations} \begin{align}
Z_I&= Q_{pI}\cR+1+c_I+\mathcal{O}(r_\perp)\,, \qquad I=1,2\,,\\
Z_3&= \left(Q_{p1}\dot{\mathbf{F}}_p\cR+\mathbf{d}_1+\mathcal{O}(r_\perp)\right)\left(Q_{p2}\dot{\mathbf{F}}_p\cR+\mathbf{d}_2+\mathcal{O}(r_\perp)\right)
+\rho_p(\lambda_p^0)\cR+c_3+1+\mathcal{O}(r_\perp) \nonumber\\	
		&= Q_{p1}Q_{p2}\abs{\dot{\mathbf{F}}_p}^2\cR^2 
+ \left[\rho_p(\lambda_p^0)+\left(Q_{p1}\mathbf{d}_2+Q_{p2}\mathbf{d}_1\right)\cdot\dot{\mathbf{F}}_p\right]\cR+\text{const.}+\mathcal{O}(r_\perp)\,.
\end{align}
\end{subequations}
Here, $\dot{\mathbf{F}}_p=\dot{\mathbf{F}}_p(\lambda_p^0)$ and $\cR$
is defined as
\begin{equation}
\cR\equiv \frac{2}{|\dot{\mathbf{F}}_p|}\ln\frac{2\abs{\dot{\mathbf{F}}_p}}{r_\perp}
\end{equation}
where $r_\perp$ is the transverse distance in $\bbR^3$ from the point
$\mathbf{F}_p(\lambda_p^0)$ on the tube. The constants $c_{I=1,2,3}$ and
$\mathbf{d}_{I=1,2}$ are defined in appendix \ref{app:CTC}\@.
Similarly, $\omega_{0,1,2}$ are expanded as 
	\begin{subequations}
	\begin{align}
	\omega_0&= (Q_{p1}+Q_{p2})\left(\dot{\mathbf{F}}_p\cdot d\mathbf{x}\right) \cR
	+\left(\mathbf{d}_1+\mathbf{d}_2\right)\cdot d\mathbf{x}
	+\mathcal{O}(r_\perp)\,,\\
	\omega_1&= Q_{p1}Q_{p2}\left(\dot{\mathbf{F}}_p\cdot d\mathbf{x}\right)\cR^2
	+\frac{\cR}{2}\left[Q_{p1}\left(\mathbf{d}_2+c_2\dot{\mathbf{F}}_p\right)+Q_{p2}\left(\mathbf{d}_1+c_1\dot{\mathbf{F}}_p\right)\right]\cdot d\mathbf{x}
	+\mathcal{O}(r_\perp)\,,\\
	\omega_2&= \frac{\cR}{2}\sum_{q(\neq p)}Q_{pq}\int d\lambda_p\,\frac{\left(\dot{\mathbf{F}}_p(\lambda_p^0)-\dot{\mathbf{F}}_q(\lambda_p)\right)\cdot d\mathbf{x}}{\abs{\mathbf{F}_p(\lambda_p^0)-\mathbf{F}_q(\lambda_p)}}+\cO(r_\perp)\,.
	\end{align}
	\end{subequations}
By plugging in the above expressions, the line element \eqref{sgi16Oct14} becomes
\begin{align}
Z^{2/3}dl^2
&= (Q_{p1}Q_{p2})^2\cR^4\abs{\dot{\mathbf{F}}_p}^2\left(d\mathbf{x}^2-\frac{\abs{\dot{\mathbf{F}}_p\cdot d\mathbf{x}}^2}{\abs{\dot{\mathbf{F}}_p}^2}\right) \nonumber\\
&\quad + (Q_{p1}Q_{p2})\cR^3\Biggl[
	\rho_p(\lambda_p^0)d\mathbf{x}^2
	+\left(\abs{\dot{\mathbf{F}}_p}^2d \mathbf{x}^2-2\abs{\dot{\mathbf{F}}_p\cdot d\mathbf{x}}^2 \right)\left(Q_{p1}\left(1+c_2\right)+Q_{p2}\left(1+c_1\right)\right)\nonumber\\
	&\qquad\qquad\qquad\qquad+\dot{\mathbf{F}}_p\cdot\left(Q_{p1}\mathbf{d}_2+Q_{p2}\mathbf{d}_1\right)d\mathbf{x}^2
	\Biggr] +\mathcal{O}(\cR^2)\,.
\end{align}
For displacement along the tube, $d\mathbf{x}\propto
\dot{\mathbf{F}}_p$, the leading $\cO(\cR^4)$ term vanishes and the
$\cO(\cR^3)$ term gives the leading contribution. If the coefficient of
the $\cO(\cR^3)$ term is negative for all $\lambda_p^0\in[0,L_p]$, the
cycle along the tube will be a CTC\@.  Conversely, for the absence of
CTCs, there must be some value of $\lambda_p^0$ for which the following
inequality is satisfied:
\begin{align} \label{nleading}
\rho_p(\lambda_p^0)\geq Q_{p1}\left(\abs{\dot{\mathbf{F}}_p}^2(1+c_2)-\dot{\mathbf{F}}_p\cdot\mathbf{d}_2\right)
+Q_{p2}\left(\abs{\dot{\mathbf{F}}_p}^2(1+c_1)-\dot{\mathbf{F}}_p\cdot\mathbf{d}_1\right)\,.
\end{align}
This can be  written more explicitly, using \eqref{c1} and \eqref{c2}, as
\begin{align} \label{finalcondition}
\rho_p(\lambda_p^0)\geq \abs{\dot{\mathbf{F}}_p(\lambda_p^0)}^2\left(Q_{p1}+Q_{p2}\right)+\sum_{q(\neq p)}Q_{pq}\int d\lambda_p\,\frac{\dot{\mathbf{F}}_p(\lambda_p^0)\cdot\left(\dot{\mathbf{F}}_p(\lambda_p^0)-\dot{\mathbf{F}}_q(\lambda_p)\right)}{\abs{\mathbf{F}_p(\lambda_p^0)-\mathbf{F}_q(\lambda_p)}}\,.
\end{align}
This is analogous to the no-CTC condition for the superthread solution
(Eq.\ (2.34) in \cite{Niehoff:2012wu}).


\subsubsection*{Charge and angular momentum}

Let us study if the solution above has the expected monopole and dipole
charges.  In the presence of Chern-Simons interaction, there are
multiple notions of charge \cite{Marolf:2000cb}, and here we choose Page
charge, which is conserved, localized, quantized, and gauge-invariant
under small gauge transformations.  Specifically, the D$p$-brane Page
charge is defined as \cite{Marolf:2000cb,
deBoer:2012ma} (see also Appendices \ref{app:convention} and \ref{app:IIAuplift})
\begin{equation}
Q_{\mathrm{D}p}^\mathrm{Page}
 =\frac{1}{(2\pi l_s)^{7-p}g_s}\int_{M^{8-p}}e^{-B_2}G
 =\frac{1}{(2\pi l_s)^{7-p}g_s}\int_{\partial M^{8-p}}e^{-B_2}C\,.
\label{Page_chg_def}
\end{equation}
Here, $M^{8-p}$ is an $(8-p)$-manifold enclosing the D$p$-brane, and
$G=\sum_{p} G_{p+1}$, $C=\sum_p C_p$ with $p$ odd (even) for type IIA
(IIB)\@.  In the integrand, we must take the part with the appropriate
rank from the polyforms $e^{-B_2}G$, $e^{-B_2}C$.  In the second
equality, we used the relation \eqref{defFT} between $G$ and $C$.

Using the definition above, we can readily calculate Page charges for
this 2-dipole solution. For example, the D4(6789)-brane charge, which is
expected to vanish, is given by
\begin{align}
Q_\mathrm{D4(6789)}^\mathrm{Page}
 &=
 \frac{1}{(2\pi l_s)^{3}g_s}\int_{S^2\times T^2_{45}}e^{-B_2}G
 =
 \frac{1}{(2\pi l_s)^{3}g_s}\int_{\partial S^2\times T^2_{45}}e^{-B_2}C
 \notag\\
 &=
 \frac{R_4 R_5}{2\pi l_s^{3}g_s}
 \int_{\partial S^2}
 \left\{
 \left[-{1\over Z_1}+{V\mu\over Z-V\mu^2}
 \left({K^1\over V}-{\mu\over Z_1}\right)
 \right]\omega
 +\xi^1\right\}\,,
\end{align}
where in the last equality we used \eqref{forms_for_Page}.  If the
surface $S^2$ is at infinity enclosing the entire profile, then the
function in the $[\cdots]$ above is single-valued.  Also, the requirement of
integrability \eqref{integrability} guarantees that $\omega$ is also
single-valued.  Therefore, the entire first term in the integrand is
single-valued and does not contribute to the integral on $\partial S^2$.
The only contribution comes from the second term, $\xi_1$.
Thus we find
\begin{align}
Q_\mathrm{D4(6789)}^\mathrm{Page}
 &=
  \frac{R_4 R_5}{2\pi l_s^{3}g_s}\int_{\partial S^2} \xi^1
 =  \frac{R_4 R_5}{2\pi l_s^{3}g_s}\int_{S^2} d\xi^1
 = - \frac{R_4 R_5}{2\pi l_s^{3}g_s}\int_{S^2} *_3 dK^1\,.
\end{align}
The integral is equal to $-4\pi$ times the coefficient of $1/r$ in the
large $r$ expansion of $K^1$.  However, $\alpha_2=\cO(1/r^2)$ and hence
$K^1=\gamma_2=\cO(1/r^2)$ and the coefficient of the $1/r$ term
vanishes.  So, we conclude that $Q_\mathrm{D4(6789)}^\mathrm{Page}=0$,
as expected.  Similarly, other Page charges are related to the
coefficient of the $1/r$ in the large $r$ expansion of the corresponding
harmonic function (see Appendix \ref{app:IIAuplift} for the expressions
for necessary RR potentials to compute the Page charge).  We find that
the non-vanishing charges are
\begin{align}
Q_\mathrm{D2(45)}^\mathrm{Page}&=Q_\mathrm{D2(67)}^\mathrm{Page}=\sum_p N_p\,,\\
Q_\mathrm{D2(89)}^\mathrm{Page}&=\sum_p N_{p3}\,,\qquad
Q_{p3}=\int_p\rho_p\,,
\end{align}
where we used \eqref{charges_d2d2d2ns5ns5}.

It is easy to check that we have appropriate monodromy for ns5($\lambda$4567) and ns5($\lambda$6780). The real part of $\tau^{1,2}$ contain $K^{1,2}$ \eqref{kahlerm} and others are all single-valued. Then we can apply same argument as \eqref{dgamma}. So we obtain
\begin{equation}
\tau^1\to\tau^1+1 \,,\qquad \tau^2\to\tau^2+1
\end{equation}
as we go around each tubes. This is proper monodromy for our system.

The angular momentum can be read off from the ADM formula
\cite{Arnowitt:1962hi}
\begin{equation}
g_{ti}=-\frac{1}{\sqrt{V(Z-V\mu^2)}}\,\omega_i=-2G_4{x^j J^{ji}\over |\xv|^3}+\cdots
\end{equation}
where $G_4$ is 4-dimensional Newton constant. By expanding $g_{ti}$ to the leading order, we obtain
\begin{equation}
-g_{ti}={x^j\over |\xv|^3}\left(
 \sum_p (Q_{p1}+Q_{p2})\int_p \dot{F}_{pi}F_{pj}
 +{1\over 4}
 \sum_{p,q}Q_{pq}\iint_{p,q} {\dot{F}_{pqi}F_{pqj}-\dot{F}_{pqj}F_{pqi}
 \over F_{pq}}
 \right)+\cO\left({1\over |\xv|^3}\right)
\end{equation}
where we used
\begin{equation}
\frac{1}{R_p}=\frac{1}{|\xv|}+\frac{\xv\cdot\mathbf{F}_p}{|\xv|^3}+\mathcal{O}\left(\frac{1}{|\xv|^3}\right)\,.
\end{equation}
Therefore the angular momentum of the 2-dipole solution is
\begin{align} \label{2dipoleangular}
 J^{ji}={1\over 4G_4}\left(
 \sum_p (Q_{p1}+Q_{p2})\int_p (\dot{F}_{pi}F_{pj}-\dot{F}_{pj}F_{pi})
 +{1\over 2}
 \sum_{p,q}Q_{pq}\iint_{p,q} {\dot{F}_{pqi}F_{pqj}-\dot{F}_{pqj}F_{pqi}
 \over F_{pq}}
 \right)\,.
\end{align}
The second term represents the contribution from the interaction between
supertubes.


\subsection{3-dipole solutions}

We can also consider a 3-dipole configuration as an extension of the
2-dipole configuration \eqref{D2D2D2ns5ns5} such as
\begin{equation} \label{3D23ns5}
\begin{array}{lclcl}
\mathrm{D2(45)}&+&\mathrm{D2(89)}&\to&\mathrm{ns5(\lambda 4589)}\\
\mathrm{D2(67)}&+&\mathrm{D2(89)}&\to&\mathrm{ns5(\lambda 6789)}\\
\mathrm{D2(45)}&+&\mathrm{D2(67)}&\to&\mathrm{ns5(\lambda 4567)}
\end{array}\,.
\end{equation}
Because there is no D6-brane, we have $V=1$.  How can we find the rest
of harmonic functions for this 3-dipole configuration, generalizing the
2-dipole solution?

First, it is natural to guess that the 3-dipole
solution has the dipole sources in all $K^{I=1,2,3}$, generalizing the
2-dipole case where $K^{I=1,2}$ had dipole sources.  Namely,
\begin{align}
\alpha^I&=\sum_p Q_{pI}\int_p \frac{\dot{\mathbf{F}}_p\cdot d\mathbf{x}}{R_p}\,,\qquad dK^I=*_3d\alpha^I\,,\qquad I=1,2,3\,.
	\end{align}
Note that the next layer of equation \eqref{Z_Ieq} to determine $Z_I$ is
quadratic in $K^I$ and therefore knows only about 2-dipole interactions.
So, we can construct $Z_I$ the same way as in the 2-dipole case, as follows:
\begin{align}\label{3dipoleharmonic}
Z_I&=1+\sum_pQ_{pI}\int_p \frac{\rho_{pI}}{R_p}\notag\\[-2ex]
&\qquad+C_{IJK}\sum_{p,q}Q_{pJ}Q_{qK}\iint_{p,q}\left[\frac{\dot{\mathbf{F}}_p\cdot\dot{\mathbf{F}}_q}{2R_pR_q}-\frac{\dot{F}_{pi}\dot{F}_{qj}(R_{pi}R_{qj}-R_{pj}R_{qi})}{F_{pq}R_pR_q(F_{pq}+R_p+R_q)}\right]\,,
\end{align}
where $I=1,2,3$ and the same shorthand notation \eqref{def_R,F,Qpq} is
used. Finally,  the last layer of equation \eqref{mu_eq} to 
determine $\mu$ is
	\begin{equation}
	\Lap\mu=\partial_iZ_I\partial_iK^I=\epsilon_{ijk}\partial_iZ_I\partial_j\alpha_k^I \,.\label{Lapmu=}
	\end{equation}
Because $Z_I$ involves 2-dipole interactions, $\mu$ involves 3-dipole
interactions.  Although we have not been able to solve this in terms of
integrals along the tubes as in the 2-dipole case (cf.\
\eqref{mu_D2D2D2ns5ns5}), we know physically that the solution must be
single-valued and therefore we can solve it by using the standard
single-valued Green function.  Namely, the solution is
	\begin{align} \label{3dipole_mu}
	\mu(\xv)=-\frac{1}{4\pi}\int d^3x'\,\frac{\partial_iZ_I\partial_iK^I(\mathbf{x'})}{|\mathbf{x}-\mathbf{x'}|}\,.
	\end{align}
In order to satisfy the integrability condition \eqref{integrability},
we have no option of adding to this a term like $\sum_p \int_p
\sigma_p/R_p$ with an arbitrary function $\sigma_p$, as we did in the
second term of \eqref{Z3_D2D2D2ns5ns5}.  In the present case, with
$V=1$, $\Lap K^I=0$, the integrability condition \eqref{integrability}
becomes
\begin{align}
0&=V\Lap M-M\Lap V
 +\frac{1}{2}\left(K^I\Lap L_I-L_I\Lap K^I\right)\notag\\
 &=\Lap M+\frac{1}{2}K^I\Lap L_I
 =\Lap \mu - \partial_i  Z_I \partial_i K^I,
\end{align}
where in the last equality we used \eqref{Zharm}, \eqref{muharm}.  This is nothing but \eqref{Lapmu=}. If we
added the term $\sum_p \int_p \sigma_p/R_p$ to the $\mu$ in
\eqref{3dipole_mu}, then the integrability condition would be violated by a
delta-function term. This is why we do not have an option of adding such a
term.  This also explains as a corollary why we do not have a term like
$\sum_p \int_p \sigma_p/R_p$ in the 2-dipole $\mu$ in
\eqref{mu_D2D2D2ns5ns5}.\footnote{In the context of the supersheet
solution \cite{Niehoff:2012wu}, (the 6D version of) this was explained from the no-CTC
condition.  }

Although it is not as explicit as the 2-dipole case, \eqref{3dipole_mu}
gives the interacting 3-dipole solution in principle.

\section{Mixed configurations}
\label{sec:mixed}

Thus far, we have studied the 4D/5D solution with codimension-2 centers.
In this section, we present a simple example in which codimension-3 and
codimension-2 centers coexist.

As the simplest codimension-2 center, let us consider the 1-dipole
configuration with the harmonic functions \eqref{harmonicD2D2},
\begin{equation}
\begin{aligned}
V&=1\,,\quad&
K^1&=0\,,& K^2&=0\,,& K^3&=\gamma\,,\\
&&L_1&=1+f_2\,,& L_2&=1+f_1\,,& L_3&=1\,,&
\quad M&=-\frac{\gamma}{2}\,,
\end{aligned}\label{harmonicD2D2(2)}
\end{equation}
where we have extracted ``1'' as compared from \eqref{f1f2_for_D2D2ns5}
and
\begin{align}
 f_1&={Q_1\over L}\int_0^L {d\lambda\over |\xv-\Fv(\lambda)|}\,,\qquad
 f_2={Q_1\over L}\int_0^L {|\dot\Fv(\lambda)|^2d\lambda\over |\xv-\Fv(\lambda)|}\,,
\end{align}
while $\gamma$ is still given by \eqref{alpha_gamma} and \eqref{alpha_def}.

We would like to add to this a codimension-3 source of the type
\eqref{codim3harm}.  Here, let us simply add a codimension-3 singularity
to \eqref{harmonicD2D2(2)} as follows:
\begin{equation}
\begin{aligned}
 V&=n_0+{n\over r}\,,\\
 K^1&=k^1_0+{k^1\over r}\,,&
 K^2&=k^2_0+{k^2\over r}\,,&
 K^3&=k^3_0+\gamma+{k^3\over r}\,,\\
 L_1&=l_1^0+f_2+{l_1\over r}\,,&
 L_2&=l_2^0+f_1+{l_2\over r}\,,&
 L_3&=l_3^0+{l_3\over r}\,,\\
 M &= m_0-{\gamma\over 2}+{m\over r}\,.
\end{aligned}
\end{equation}
For these harmonic functions, the integrability condition
\eqref{integrability} becomes
\begin{align}
 0
 &=-4\pi\delta(\xv)
 \left[n_0 m-m_0 n +{1\over 2}(k_0^I l_I - l_I^0 k^I)
 -{1\over 2}\Bigl(k^1 f_2(\xv=0)+k^2 f_1(\xv=0)\Bigr)\right]
 \notag\\
 &\quad
 -2\pi\gamma \,\delta(\xv)(n+l_3)\notag\\
 &\quad
 +{1\over 2}\left[
 \left(k_0^2+{k^2\over r}\right)\Lap f_1
 + \left(k_0^1+{k^1\over r}\right)\Lap f_2
 \right]\,.
\end{align}
The three lines on the right hand side are of different nature and must
vanish separately.  So,
\begin{subequations} 
 \begin{align}
 0&=n_0 m-m_0 n +{1\over 2}(k_0^I l_I - l_I^0 k^I)
 -{1\over 2}{Q\over L}\int_0^L{d\lambda}{k^1 |\dot{\Fv}(\lambda)|^2+k^2
 \over |\Fv(\lambda)|}\,,
  \label{jghm2Mar15}\\
 0&=n+l_3\,,\label{hmg4Mar15}\\
 0&=
 k_0^2+{k^2\over |\Fv(\lambda)|}
 +|\dot{\Fv}(\lambda)|^2
 \left(
 k_0^1+{k^1\over |\Fv(\lambda)|}
  \right)\qquad \text{for each value of $\lambda$.}
 \label{jgla2Mar15}
 \end{align}
\end{subequations}
The first equation \eqref{jghm2Mar15} says that the total force exerted
by the tube on the $r=0$ brane must vanish.  This is a single equation
and easy to satisfy.  The second equation is also easy to satisfy.  On
the other hand, the third equation \eqref{jgla2Mar15} says that the
force exerted by the $r=0$ brane on every point of the tube must
vanish, and gives the most stringent condition.  Let us investigate this
last condition in detail.

Note that, if the asymptotic moduli $k_0^1,k_0^2$ vanished, then the
distance between the tube and the codimension-3 brane, $|\Fv(\lambda)|$, would
disappear from the condition \eqref{jgla2Mar15}, and we have
\begin{align}
 0&=
 {k^2} +|\dot{\Fv}(\lambda)|^2{k^1}\,.\label{jijo2Mar15}
\end{align}
Because $|\dot{\Fv}(\lambda)|^2$ is the ratio of the D2(67) and D2(45)
charge densities carried by the tube while $k^1,k^2$ are the D4(6789),
D4(4589) charges of the $r=0$ brane, Eq.\ \eqref{jijo2Mar15} would mean
that the tube must have, at every point along it, charge density that
would be mutually supersymmetric with the $r=0$ brane in flat space.
This can of course happen only if the total charge of the tube is
mutually supersymmetric with the $r=0$ brane. In this case, the distance
between the two objects is arbitrary, implying that they are not bound.

On the other hand, if the asymptotic moduli $k_0^1,k_0^2$ are
non-vanishing, the tube does not have charge density that would be
mutually BPS with the $r=0$ brane in flat space, and
the configuration represents a  true bound state.
The condition \eqref{jgla2Mar15} gives
\begin{align}
 |\dot{\Fv}(\lambda)|^2
 = - {k_0^2 |\Fv(\lambda)| + k^2 \over  k_0^1 |\Fv(\lambda)| + k^1 }\,.
\end{align}
Because $\Fv(\lambda)$ is a vector with three components, this
differential equation leaves the orientation of $\dot{\Fv}(\lambda)$
undetermined.  Therefore, the tube profile can wiggle depending on
\emph{two} functions of one variable.  We expect that this remains true
for more general configurations with both codimension-2 and
codimension-3 centers: each codimension-2 center has a profile depending
on two functions of one variable, so that the force from other centers
vanishes at each point along the tube.

\section{Discussion}
\label{sec:discussion}

In this paper, we studied the BPS configurations of the brane system in
string theory in the framework of 5D supergravity.  In the literature,
multi-center configurations of codimension-3 branes have been
extensively studied.  However, we pointed out that these codimension-3
branes can polarize into codimension-2 ones by the supertube effect and
hence multi-center configurations involving codimension-2 branes along
arbitrary curves must also be included if we want to capture the full
configuration space of the system.  Codimension-2 branes can be exotic,
and the solution containing them can represent non-geometric spacetime.

Therefore, the most general configuration is a multi-center
configuration including both codimension-3 branes and codimension-2
ones.  In the framework of the 4D/5D solution, such configurations are
described by harmonic functions with codimension-3 and codimension-2
singularities in $\bbR^3$. In this paper, we provided some simple
examples of such solutions, hoping that they serve as a guide for
constructing general solutions.

The solutions with codimension-2 centers have various possible
applications and implications, some of them already mentioned in the
Introduction.  Here let us discuss their relevance to the fuzzball
proposal for black holes 
\cite{Mathur:2005zp,Bena:2007kg,Skenderis:2008qn,Balasubramanian:2008da,Chowdhury:2010ct}
and the microstate geometry program.

Smooth 4D/5D solutions with codimension-3 centers have been put forward
as possible microstates for the 3- and 4-charge black holes
\cite{Bena:2005va, Berglund:2005vb}.  However, the entropy represented
by these solutions have been estimated \cite{deBoer:2009un, Bena:2010gg}
to be parametrically smaller than the entropy of the corresponding black
hole.  In particular, for the 3-charge black hole, Ref.\
\cite{Bena:2010gg} considered placing a probe supertube in the scaling
geometry \cite{Bena:2006kb, Bena:2007qc} and estimated the associated
entropy to be $\sim Q^{5/4}$ whereas the desired black hole entropy is
$\sim Q^{3/2}$, where $Q\sim Q_{1,2,3}$ is the charge of the black hole.
In our setup, a supertube in a scaling geometry corresponds to a
configuration with codimension-3 centers as well as a codimension-2 one.
It may be possible to make their estimate more precise by including
backreaction using our setup.

Another issue with identifying smooth 4D/5D solutions with codimension-3
centers with black hole microstates concerns the pure Higgs branch.
Ref.~\cite{Bena:2012hf} (see also \cite{Lee:2012sc}) studied quiver quantum mechanics describing
3-center solutions  and showed that most entropy of the
system comes from zero-angular momentum states in what they call the
pure Higgs branch.  On the other hand, the multi-center solutions with
codimension-3 centers are naturally identified with states in the
Coulomb branch of the quiver quantum mechanics. This is because the
codimension-3 solutions are characterized by the position of the centers,
which corresponds to the adjoint vev in the quiver quantum mechanics.
Therefore, these solutions do not seem to correspond to typical
microstates of the system.  In contrast, a codimension-2 center
has a finite-sized \emph{profile}, as a result of two branes getting
bound together and puffing up by the supertube effect.  In the quiver
quantum mechanics, this has a natural interpretation as a Higgs branch
state, with a finite vev for the bifundamental matter connecting two
centers or nodes.  Therefore, it is very interesting to understand the
relation between the codimension-2 configurations in gravity and states
in quiver quantum mechanics to elucidate the role of codimension-2
centers in black hole microphysics.

We have focused on codimension-2 centers in this paper but, of course,
we could consider objects with still lower codimensions, namely one and
zero.  A codimension-1 center is a membrane in $\bbR^3$ and is a
4D/5D-solution realization of the ``superstrata'' recently proposed as
possible microstates \cite{deBoer:2010ud, deBoer:2012ma, Bena:2011uw,
Bena:2014qxa}.  It is interesting to study if the setup of the 4D/5D
solution sheds new light on superstrata or makes their construction and
analysis easier.  Codimension-1 and codimension-0 branes are generally
more non-geometric than the codimension-2 ones \cite{Hassler:2013wsa, Andriot:2014uda},
and studying them in the context of the 4D/5D solution is an interesting
subject.

Explicit construction of a solution with codimension-2 centers with
general charge, position and profile is technically a challenging
problem.
In subsection \ref{ss:2-dip}, we discussed how to solve the BPS
equations of subsection \ref{4D5Dsol} for a 2-dipole supertube. As
mentioned there, when solving the BPS equations, there are multiple
solutions differing in the monodromy properties. We must construct them
and choose from them the physically appropriate one expected from the
dipole charges produced by supertube transitions.  This is in some sense
similar to (but more complicated than) the problem of finding solutions
of F-theory with various monodromies around 7-branes
\cite{Greene:1989ya, Vafa:1996xn, Bergshoeff:2006jj} and is a
non-trivial task.  In particular, in the presence of non-trivial
harmonic function $V$, which corresponds to having D6-branes, solving
Eq.\ \eqref{Z_Ieq} is itself a challenging problem. We leave this for
future research.

To conclude, the solutions involving codimension-2 provide interesting
new directions of research, and studying them is bound to reveal richer
physics of brane systems than was found in codimension-3 solutions.  We
hope to report on the progress in such research in near future.

\section*{Acknowledgments}

We would like to thank Iosif Bena, Jan de Boer, Stefano Giusto, Daniel
Mayerson, Ben Niehoff, Rodolfo Russo, Orestis Vasilakis, and Nick Warner
for valuable discussions.  The work of MS was supported in part by
Grant-in-Aid for Young Scientists (B) 24740159 from the Japan Society
for the Promotion of Science (JSPS)\@.

\appendix
\section{Convention}
\label{app:convention}

The reduction formulas for the 11D metric and 3-form potential to type
IIA supergravity in 10D are
\begin{equation}
\label{M-IIA}
\begin{split}
ds_{11}^2&=e^{-\frac{2}{3}\Phi}ds_\mathrm{10,str}^2+e^{\frac{4}{3}\Phi}\left(dx^{11}+C_1\right)^2\,,\\
\cA_3&=C_3+B_2\wedge dx^{11}\,.
\end{split}
\end{equation}

The relation between the gauge-invariant RR field strength $G_{p+2}$ and
the RR potential $C_{p+1}$ is
\begin{align} \label{defFT1}
 G_{p+2}=dC_{p+1}-H_3\wedge C_{p-1}\,,
\end{align}
where $H_3=dB_2$.
The higher forms
$G_6,G_8$ are related to $G_4,G_2$ by
\begin{align}
G_6=*G_4\,,\qquad G_8=-*G_2\,.
\end{align}
If we define the polyforms $G=\sum_p G_{p+1}$, $C=\sum_p C_{p}$ with $p$
odd (even) for type IIA (IIB), the relation \eqref{defFT1} can be written
more concisely as
\begin{align} \label{defFT}
 G=dC-H_3\wedge C=e^{B_2}d(e^{-B_2}C)\,.
\end{align}

We define the Hodge dual of a $p$-form $\omega$ in $d$ dimensions as
\begin{align}
  (* \, \omega)_{i_1 \cdots i_{d-p}} &= \frac{1}{p!}\,\epsilon_{i_1
  \cdots i_{d-p}}{}^{ j_1 \cdots j_p} \omega_{j_1 \cdots j_p}\,,
 \\
   *(dx^{j_1}\wedge\cdots\wedge dx^{j_p})
 &= \frac{1}{(d-p)!}\,
 dx^{i_1}\wedge\cdots\wedge dx^{i_{d-p}}
 \epsilon_{i_1 \cdots i_{d-p}}{}^{ j_1 \cdots j_p} \,,
\end{align}
with
\begin{align}
 \epsilon_{01\dots (d-1)}=-\sqrt{-g}\,,\qquad
 \epsilon^{01\dots (d-1)}=+{1\over \sqrt{-g}}\,.
\end{align}

\section{Monodromic harmonic function}
\label{app:alpha}

Here, we show that if the harmonic function $\gamma$ has the monodromy
\eqref{dgamma} independent of the cycle $c$, then it is given in terms
of the 1-form $\alpha$ by \eqref{alpha_gamma} and \eqref{alpha_def}.

Harmonicity of $\gamma$ means that $d(*_3d\gamma)=0$, which implies that
$*_3 d\gamma$ is closed and can be written in terms of a 1-form $\alpha$
as $*_3 d\gamma=d\alpha$ at least locally. Because $\alpha$ has the
gauge ambiguity $\alpha \to \alpha+d\Lambda$ where $\Lambda$ is a
scalar, we can impose the ``Lorenz gauge'' $\partial_i \alpha_i=0$.  In
this gauge, the monodromy of $\gamma$ can be expressed as
\begin{align}
 \Delta \gamma 
 =\int _c d\gamma 
 =\int _c *_3d\alpha
 =\int _D d*_3d\alpha
 =-\int _D \Lap \alpha_i\, {1\over 2}\epsilon_{ijk}dx^j \wedge dx^k
 =-\int _D \Lap \alpha_i\, n_i\, d^2\! A\,,\label{ncvo1May15}
\end{align}
where $D$ is a 2-surface with $\partial D=c$, $n_i$ is the unit normal
to $D$, and $d^2\!A$ is the area element of $D$.  In order for the
monodromy $\Delta \gamma$ not to change even if we homotopically
deform the cycle $c$, the quantity $\Lap \alpha$ can only have
delta-function source along the profile $\xv=\Fv(\lambda)$.  Therefore,
it must be that
\begin{align}
\alpha_i(\xv) = {1\over L}\int_0^L {v_i(\lambda)\over |\xv-\Fv(\lambda)|}d\lambda
\end{align}
where $v_i(\lambda)$ are some functions. This gives
\begin{align}
\Lap\alpha_i(\xv)
 = -{4\pi \over L}\int_0^L v_i(\lambda)\, \delta^2(\xv-\Fv(\lambda))\,d\lambda\,.
\end{align}
Namely, $\alpha_i$ has delta-function source distributed along the
profile with (vectorial) density $v_i$.  Then \eqref{ncvo1May15}
is proportional to
\begin{align}
 v_i n_i \times {1\over \cos\theta}
 \times {1\over |\dot{\Fv}|}\,,\label{ndxb1May15}
\end{align}
where $\theta$ is the angle between $n_i$ and the unit tangent to the
profile, $t_i$.  The second factor takes into account the fact that the
curve does not necessarily perpendicularly intersect with $D$, and the
third factor takes into account the ``speed'' of the parametrization
$\lambda$.  Because $\cos\theta=t_j n_j$ and
$t_j=\dot{F}_j/|\dot{\Fv}|$, the quantity \eqref{ndxb1May15} is equal to
\begin{align}
 {v_i n_i \over t_j n_j}\,.\label{neig1May15}
\end{align}
Given $c$, there are infinitely many choices for $D$ which
can intersect the profile at any point at any angle.  So, if
\eqref{neig1May15} is to be independent of the choice of $D$, the only
possibility is $v_i\propto \dot{F}_i$.  This means that 
$\alpha$ is given by \eqref{alpha_def}.

\section{Superthread} \label{app:superthread}

In this Appendix, we briefly review the superthread solution which was
used in subsection \ref{ss:2-dip} to derive the 2-dipole solution. The
superthread solution was originally obtained in \cite{Niehoff:2012wu} as a BPS solution in 6D
supergravity \cite{Bena:2011dd}.

The metric for the superthread is 
\begin{equation}
	ds_6^2=2(Z_1Z_2)^{-1/2}dv\left(du+k+\frac{1}{2}\mathcal{F}\,dv\right)-(Z_1Z_2)^{1/2}ds_4^2\label{metric_superthread}
\end{equation}
where the base space is flat $\bbR^4$ with metric
$ds_4^2=\delta_{ij}dx^idx^j$ ($i=1,\cdots,4$).  We denote the coordinates
of the $\bbR^4$ by $\vec{x}=(x^1,x^2,x^3,x^4)$.  All quantities that
appear in the metric are independent of the coordinate $u$.  The scalars
$Z_I$, $I=1,2$ are harmonic functions in $\mathbb{R}^4$ and are given by
\begin{equation}
Z_I=1+\sum_p\frac{Q_{pI}}{R_p^2}\,,\label{Z_I_superthread}
\end{equation}
where
\begin{equation}
R_p\equiv\abs{\vec{x}-\vec{F}^{(p)}(v)}
\end{equation}
and $\vec{F}^{(p)}(v)\in\bbR^4$ is the profile of the supertube.  Note
that we use this $\bbR^4$ version of $R_p$ only in this appendix ($R_p$
in the main text is defined for $\bbR^3$ as in \eqref{def_R,F,Qpq}).
The 6D solution also involve self-dual field strengths
\begin{equation}
\Theta^I=*_4\Theta^I\,,\quad I=1,2\,,
\end{equation}
which are related to $Z_I$ by the following equation:
\begin{equation}
d\Theta^I=|\epsilon^{IJ}|\, {*_4d\dot{Z}_J}\,.
\end{equation}
Here $\dot{~~}$ means the $v$-derivative and $d$ is the exterior
derivative with respect to the $\bbR^4$.  For $Z_I$ given in
\eqref{Z_I_superthread}, this equation can be solved by
\begin{equation}
\Theta^I=(1+*_4)d\left(|\epsilon^{IJ}|\sum_pQ_{pJ}\frac{\dot{\vec{F}}^{(p)}\cdot d\vec{x}}{R_p^2}\right)\,.
\end{equation}
The 1-form $k$ appearing in the metric \eqref{metric_superthread} satisfies 
the relation
\begin{equation}
(1+*_4)dk=Z_I\Theta^I\,.
\end{equation}
The solution to this equation is
\begin{subequations} \label{ihxn5Aug11}
\begin{align}
k&=k_0+k_1+k_2\,,\\
k_0&=\sum_{I=1,2}\sum_{p}{Q_{pI}\dot{\vec{F}}^{(p)}\cdot d\vec{x}\over R_p^2}\,,\\
k_1&={1\over 2}\sum_{p,q}Q_{pq}{\dot{\vec{F}}^{(q)}\cdot d\vec{x}\over  R_p^2R_q^2}
	={1\over 4}\sum_{p,q}Q_{pq}{(\dot{\vec{F}}^{(p)}+\dot{\vec{F}}^{(q)})\cdot d\vec{x}\over  R_p^2R_q^2}\,,\\
k_2&={1\over 4}\sum_{p,q}Q_{pq}{\dot{F}^{(p)}_i - \dot{F}^{(q)}_i\over \abs{\vec{F}^{(p)}-\vec{F}^{(q)}}^2}
\left[\left({1\over R_p^2}-{1\over R_q^2} \right)dx^i-{2\over R_p^2 R_q^2}\cA_{ij}^{(p,q)}dx^j \right]\,,
\end{align}
\end{subequations}
where we defined
\begin{equation}
Q_{pq}\equiv Q_{p1}Q_{q2}+Q_{q1}Q_{p2}\,.
\end{equation}
With this $k$, the scalar field $\mathcal{F}$ can be obtained by
solving the equation
\begin{equation}
-*_4d*_4d\mathcal{F}=*_4(\Theta^1\wedge\Theta^2)+2\dot{Z_1}\dot{Z_2}\,.
\end{equation}
This can be solved by
\begin{equation}
\cF=-1-\sum_p {\rho_p\over R_p^2}-\sum_{p,q}Q_{pq}
\left[\frac{\dot{\vec{F}}^{(p)}\cdot\dot{\vec{F}}^{(q)}}{2R_p^2R_q^2}-{\dot{F}^{(p)}_i\dot{F}^{(q)}_j \cA^{(p,q)}_{ij}\over R_p^2R_q^2\abs{\vec{F}^{(p)}-\vec{F}^{(q)}}^2}\right]\,,
\end{equation}
where
\begin{equation}
\cA_{ij}^{(p,q)}\equiv R_i^{(p)}R_j^{(q)}- R_j^{(p)}R_i^{(q)}-\epsilon^{ijkl}R_k^{(p)}R_l^{(q)}\,.
\end{equation}

After smearing out the above solution along $x^4$ and $v$
directions\footnote{The smearing along $v$ is similar to that in
\cite{Lunin:2001fv}.} and identifying quantities as stated in
\cite{Niehoff:2013kia}, we can reinterpret the quantities above
($Z_I,\Theta^I,k,\cF$) in terms of the harmonic functions appearing in
the 4D/5D solution.  Specifically, we obtain $V=1$, $K^3=\Theta^3=0$,
$\mathcal{F}=-Z_3$. All other quantities can be read off from the
relations \eqref{Zharm}, \eqref{k_decompose}, \eqref{muharm}, and
\eqref{omega_eq}.

\section{Near-tube expansions} \label{app:CTC}

In this appendix, we carry out the near-tube expansions of quantities
that are used in the no-CTC analysis in the main text.  To avoid
clutter, we suppress the subscript $p$ from the quantities such as
$\Fv_p$ and $\lambda_p$ associated with the $p$-th tube.

We want to evaluate the near-tube limit of quantities such as
\begin{align}
I(\xv)\equiv \int {d\lambda \over |\xv-\Fv(\lambda)|}\,.\label{jfkj16Feb15}
\end{align}
Consider a point $\xv$ very close to the tube.  Near the point $\xv$,
the tube can be thought of as a straight line. Let us take a cylindrical
coordinate system $(r_\perp,\theta,z)$ in which the point $\xv$ is at
$\theta=z=0$.  Also, let the point $r_\perp=z=0$ on the curve (which is
now a line) be $\Fv(\lambda^0)$ where $\lambda^0$ is the value of
the parameter corresponding to that point. Both the points $\xv$ and
$\Fv(\lambda^0)$ are in the $z=0$ plane. Then, by approximating the
curve by a straight line there,
	\begin{equation}
	\abs{\mathbf{x}-\mathbf{F}(\lambda)}\approx\sqrt{r_\perp^2+\abs{\dot{\mathbf{F}}(\lambda^0)}^2(\lambda-\lambda^0)^2}
	\end{equation}
where $r_\perp$ is the radial distance from the curve.  For very small
$r_\perp$, most contribution to the integral \eqref{jfkj16Feb15} comes
from very small $|\lambda-\lambda^0|$.  So, let us introduce a small
cutoff $\epsilon>0$ and divide the integral as
\begin{align}
 \int d\lambda
 &=
 \int_{\lambda^0-\epsilon}^{\lambda^0+\epsilon}d\lambda
 +\int^{\lambda^0-\epsilon}d\lambda
 +\int_{\lambda^0+\epsilon}d\lambda\label{sbe17Feb15}
 \\
 &\equiv
 \int_{\lambda^0-\epsilon}^{\lambda^0+\epsilon}d\lambda
 +P_\epsilon\!\! \int d\lambda
 \label{sbx17Feb15}
\end{align}
where $P_\epsilon\!\int$ means to exclude the interval
$[\lambda^0-\epsilon,\lambda^0+\epsilon]$ from the integral. 
We take the following limit:
\begin{align}
 r_\perp\to 0\,,\qquad \epsilon\to 0\,,\qquad\text{with}\qquad
 {r_\perp\over \epsilon}\to 0\,.\label{jglw16Feb15}
\end{align}
We take $\epsilon\to 0$ so that the curve for
$\lambda\in[\lambda^0-\epsilon,\lambda^0+\epsilon]$ can be regarded as a
straight line.  Because we are very close to the straight line, we must
take $r_\perp\to 0$, ${r_\perp\over \epsilon}\to 0$.

In this limit, the first term in \eqref{jglw16Feb15} is evaluated as
\begin{align}
 \int_{\lambda^0-\epsilon}^{\lambda^0+\epsilon}
{d\lambda \over |\xv-\Fv(\lambda)|}
 &\approx 
  \int_{\lambda^0-\epsilon}^{\lambda^0+\epsilon}
{d\lambda\over\sqrt{r_\perp^2+|\dot{\Fv}|^2(\lambda-\lambda^0)^2}}
 \notag\\
 &\approx
  {1\over |\dot{\Fv}|}\int_{-|\dot{\Fv}|\epsilon}^{|\dot{\Fv}|\epsilon}
{d\lambda'\over\sqrt{r_\perp^2+\lambda'^2}}
 \approx
 {2\over |\dot{\Fv}|}
 \log\biggl({2\epsilon |\dot{\Fv}|\over r_\perp}\biggr)
\end{align}
where $\dot{\Fv}\equiv \dot{\Fv}(\lambda^0)$ and
$|\dot{\Fv}|(\lambda-\lambda^0)\equiv \lambda'$. This diverges as
$\epsilon/r_\perp\to \infty$ because the contribution from an infinite
straight line is infinite. However, of course, the actual curve is
finite and closed, and the integral must be finite. In other words, in
the full integral \eqref{sbx17Feb15}, $\epsilon$-dependence must cancel
out.  Therefore, we must be able to split $I(\xv)$ as follows:
\begin{equation}
I(\mathbf{x})=\frac{2}{|\dot{\mathbf{F}}|}\ln\frac{2\abs{\dot{\mathbf{F}}}}{r_\perp}
+
\lim_{\epsilon\to 0}\left[
P_\epsilon\!\!\int\frac{d\lambda}{|\mathbf{F}(\lambda)-\mathbf{F}(\lambda^0)|}
+\frac{2}{|\dot{\mathbf{F}}|}\ln\epsilon
\right]
+\mathcal{O}(r_\perp)\,,\label{tia17Feb15}
\end{equation}
where $[\dots]$ is finite in the $\epsilon\to 0$ limit.
Indeed, the second term in \eqref{sbe17Feb15} is
\begin{align}
 \int^{\lambda^0-\epsilon}
\frac{d\lambda}{|\mathbf{x}-\mathbf{F}(\lambda)|}
 &\approx 
 \int^{\lambda^0-\epsilon}
\frac{d\lambda}{|\mathbf{F}(\lambda^0)-\mathbf{F}(\lambda)|}
\label{jsxb11May15}
\end{align}
and includes a divergent contribution from near the upper bound of the
integral, $\lambda=\lambda^0-\epsilon$.  The diverging contribution can
be evaluated as
\begin{align}
 \eqref{jsxb11May15}&\approx 
 \frac{1}{|\dot{\mathbf{F}}|}
 \int^{-\epsilon}
 \frac{d\lambda'}{|\lambda'|}
 \approx 
 -\frac{1}{|\dot{\mathbf{F}}|}\ln\epsilon\,.
\end{align}
We get an identical contribution from the third term in \eqref{sbe17Feb15}. These divergences precisely  cancel the second
term in $[\dots]$ of \eqref{tia17Feb15}. 

So, for example, as we  approach the point $\Fv_p(\lambda_p^0)$
on the $p$-th tube, the behavior of the integral appearing in $Z_{I=1,2}$ of
\eqref{2tubeharmonic_L1L2} is
\begin{align} 
\sum_q Q_{qI}\int_q {1\over R_q}
=
\sum_q {Q_{qI}\over L_q}\int \frac{d\lambda_q}{\abs{\mathbf{x}-\mathbf{F}_q(\lambda_q)}}
 ={Q_{pI}\over L_p}\cR+c_{I}+\mathcal{O}(r_\perp)
\end{align}
(see \eqref{def_int_p} for the first equality) where $c_{I=1,2}$ is defined by
\begin{equation} \label{c1}
c_{I}\equiv {Q_{pI}\over L_p}\lim_{\epsilon\to 0}\left[
 P_\epsilon\!\!\int\frac{d\lambda_p}
 {|\mathbf{F}_p(\lambda_p^0)-\mathbf{F}_p(\lambda_p)|}
 +\frac{2}{|\dot{\mathbf{F}}_p|}\ln\epsilon
 \right]
 +\sum_{q(\neq p)}{Q_{qI}\over L_q}
 \int \frac{d\lambda_q}
 {\abs{\mathbf{F}_p(\lambda_p^0)-\mathbf{F}_q(\lambda_q)}}
\end{equation}
and is independent of $r_\perp$. We also defined
\begin{align}
\cR\equiv \frac{2}{|\dot{\mathbf{F}}_p|}\ln\frac{2\abs{\dot{\mathbf{F}}_p}}{r_\perp}\,.
\end{align}

Using the same argument, we can also derive the behavior of the
integrals appearing in $\omega$ and $Z_3$ as follows:
\begin{align}
\sum_q Q_{qI}\int_q \frac{\dot{\mathbf{F}}_q(\lambda_q)}{R_q(\lambda_q)}
 &=
\sum_q {Q_{qI}\over L_q}
 \int  \frac{\dot{\mathbf{F}}_q(\lambda_q)\,d\lambda_q}{\abs{\mathbf{x}-\mathbf{F}_q(\lambda_q)}}
=
{Q_{pI}\over L_p}\dot{\mathbf{F}}_p(\lambda_p^0)\cR
+\mathbf{d}_I
+\mathcal{O}(r_\perp)\,,
\\
\sum_q \int_q \frac{\rho_q}{R_q}
 &=
\sum_q {1\over L_q}
 \int \frac{\rho_q(\lambda_q)\, d\lambda_q}{\abs{\mathbf{x}-\mathbf{F}_q(\lambda_q)}}
=
{1\over L_p}\rho_p(\lambda_p^0)\cR
+c_3
+\mathcal{O}(r_\perp)\,,
\end{align}
where
\begin{align} \label{c2}
\mathbf{d}_I
&\equiv
{Q_{pI}\over L_p}\lim_{\epsilon\to 0}\left[
P_\epsilon\!\!\int
\frac{\dot{\mathbf{F}}_p(\lambda_p)\,d\lambda_p}{|\mathbf{F}_p(\lambda_p^0)-\mathbf{F}_p(\lambda_p)|}
+
\frac{2\dot{\mathbf{F}}_p(\lambda_p^0)}{|\dot{\mathbf{F}}_p|}\ln\epsilon\right]
+\sum_{q(\neq p)} {Q_{qI}\over L_q}
\int \frac{\dot{\mathbf{F}}_q(\lambda_q)\,d\lambda_q}
{\abs{\mathbf{F}_p(\lambda_p^0)-\mathbf{F}_q(\lambda_q)}}\,,
\\
c_3
&\equiv
{1\over L_p}\lim_{\epsilon\to 0}\left[
P_\epsilon\!\!\int
\frac{\rho_p(\lambda_p)\,d\lambda_p}{|\mathbf{F}_p(\lambda_p^0)-\mathbf{F}_p(\lambda_p)|}
+
\frac{2\rho_p(\lambda_p^0)}{|\dot{\mathbf{F}}_p|}\ln\epsilon\right]
+\sum_{q(\neq p)} {1\over L_q}
\int \frac{\rho_q(\lambda_q)\,d\lambda_q}
{\abs{\mathbf{F}_p(\lambda_p^0)-\mathbf{F}_q(\lambda_q)}}\,.
\end{align}

\section{The type IIA uplift and Page charges}
\label{app:IIAuplift}

The type IIA uplift of the 4D/5D solution is, including
higher RR potentials (cf.\ \eqref{IIAfield}),
\begin{align}
\begin{split}
 ds_\text{IIA,10}^2&=-{1\over \sqrt{V\Sigma}}\,\dtt^2
 +\sqrt{V \Sigma}\,dx_{123}^2
+
 \sqrt{\Sigma\over V}\left(Z_1^{-1}dx_{45}^2+Z_2^{-1}dx_{67}^2+Z_3^{-1}dx_{89}^2\right),\\
 e^{2\Phi}&={\Sigma^{3/2}\over V^{3/2} Z}\,,\qquad
 B_2=\Lambda^I J_I\,,\\
 C_1&=-{V\mu\over \Sigma}\,\dtt+A\,,\qquad
 C_3=\left(-Z_I^{-1}\dtt+\Lambda^I A+\xi^I\right)\wedge J_I\,,\\
 C_5&=\left({\mu\over Z_2 Z_3}\dtt+\Lambda^2 \Lambda^3 A +\Lambda^2 \xi^3 + \Lambda^3 \xi^2 +\zeta_1\right)\wedge J_2 \wedge J_3+\text{(cyclic)}\,,\\
 C_7&=\left({\Sigma\over ZV}\dtt+\Lambda^1 \Lambda^2 \Lambda^3 A
 +\Lambda^1 \Lambda^2 \xi^3+\Lambda^2\Lambda^3 \xi^1+\Lambda^3\Lambda^1 \xi^2
  +\Lambda^I \zeta_I
 +W\right)
 \wedge J_1 \wedge J_2 \wedge J_3\,,
\end{split}
\end{align}
where
\begin{align}
 \dtt&\equiv dt+\omega\,,\qquad
 \Sigma\equiv Z-V\mu^2\,,\qquad
 \Lambda^I\equiv V^{-1}K^I-Z_I^{-1}\mu\,,
\end{align}
and the 1-forms $(A,\xi^I,\zeta_I,W)$ are related to the
harmonic functions $(V,K^I,L_I,M)$ by
\begin{align}
 dA=*_3 dV\,,\qquad
 d\xi^I=-*_3 dK^I\,,\qquad
 d\zeta_I=-*_3 dL_I\,,\qquad
 dW=-2*_3 dM\,.
\end{align}

The expressions for forms that are useful for computing the Page charge
\eqref{Page_chg_def}
are
\begin{align}
 \label{forms_for_Page}
\begin{split}
 e^{-B_2}C|_1
 &=-{V\mu\over \Sigma}\,\dtt+A\,,\\
 e^{-B_2}C|_3
 &=\left[\left(-{1\over Z_1}+{V\mu \Lambda^1\over \Sigma}\right)\dtt+\xi^1\right]\wedge J_1+\text{(cyclic)}\,,\\
 e^{-B_2}C|_5
 &=
 \left[
 \left({Z_1 \mu\over Z}+{\Lambda^2\over Z_3}+{\Lambda^3\over Z_2}
 -{V\mu \Lambda^2 \Lambda^3\over \Sigma}\right)\dtt+\zeta_1
 \right]\wedge J_2 \wedge J_3 +\text{(cyclic)}\,, \\
 e^{-B_2}C|_7
 &=
 \left[\left(
 {\Sigma\over ZV}-{\mu\over Z}\Lambda^I Z_I
 -{\Lambda^2 \Lambda^3\over Z_1}
 -{\Lambda^3 \Lambda^1\over Z_2}
 -{\Lambda^1 \Lambda^2\over Z_3}
 +{V\mu \Lambda^1 \Lambda^2 \Lambda^3 \over \Sigma}
 \right)\dtt
 +W\right]\wedge J_1\wedge J_2 \wedge J_3\,,
\end{split}
 \end{align}
where $X|_p$ means  the $p$-form part of the polyform $X$.



\begin{thebibliography}{10}

\bibitem{Greene:1989ya}
B.~R. Greene, A.~D. Shapere, C.~Vafa, and S.-T. Yau, ``{Stringy Cosmic Strings
  and Noncompact Calabi-Yau Manifolds},''
\href{http://dx.doi.org/10.1016/0550-3213(90)90248-C}{{\em Nucl.Phys.}
  {\bfseries B337} (1990) 1}.

\bibitem{Vafa:1996xn}
C.~Vafa, ``{Evidence for F-Theory},''
  \href{http://dx.doi.org/10.1016/0550-3213(96)00172-1}{{\em Nucl. Phys.}
  {\bfseries B469} (1996) 403--418},
\href{http://arxiv.org/abs/hep-th/9602022}{{\ttfamily arXiv:hep-th/9602022
  [hep-th]}}.

\bibitem{deBoer:2010ud}
J.~de~Boer and M.~Shigemori, ``{Exotic branes and non-geometric backgrounds},''
  \href{http://dx.doi.org/10.1103/PhysRevLett.104.251603}{{\em Phys.Rev.Lett.}
  {\bfseries 104} (2010) 251603},
\href{http://arxiv.org/abs/1004.2521}{{\ttfamily arXiv:1004.2521 [hep-th]}}.

\bibitem{deBoer:2012ma}
J.~de~Boer and M.~Shigemori, ``{Exotic Branes in String Theory},''
  \href{http://dx.doi.org/10.1016/j.physrep.2013.07.003}{{\em Phys.Rept.}
  {\bfseries 532} (2013) 65--118},
\href{http://arxiv.org/abs/1209.6056}{{\ttfamily arXiv:1209.6056 [hep-th]}}.

\bibitem{Mateos:2001qs}
D.~Mateos and P.~K. Townsend, ``{Supertubes},''
  \href{http://dx.doi.org/10.1103/PhysRevLett.87.011602}{{\em Phys.Rev.Lett.}
  {\bfseries 87} (2001) 011602},
\href{http://arxiv.org/abs/hep-th/0103030}{{\ttfamily arXiv:hep-th/0103030
  [hep-th]}}.

\bibitem{Gauntlett:2002nw}
J.~P. Gauntlett, J.~B. Gutowski, C.~M. Hull, S.~Pakis, and H.~S. Reall, ``{All
  supersymmetric solutions of minimal supergravity in five dimensions},''
  \href{http://dx.doi.org/10.1088/0264-9381/20/21/005}{{\em Class.Quant.Grav.}
  {\bfseries 20} (2003) 4587--4634},
\href{http://arxiv.org/abs/hep-th/0209114}{{\ttfamily arXiv:hep-th/0209114
  [hep-th]}}.

\bibitem{Bena:2004de}
I.~Bena and N.~P. Warner, ``{One Ring to Rule Them All ... and in the Darkness
  Bind Them?},'' \href{http://dx.doi.org/10.4310/ATMP.2005.v9.n5.a1}{{\em
  Adv.Theor.Math.Phys.} {\bfseries 9} (2005) 667--701},
\href{http://arxiv.org/abs/hep-th/0408106}{{\ttfamily arXiv:hep-th/0408106
  [hep-th]}}.

\bibitem{Gutowski:2004yv}
J.~B. Gutowski and H.~S. Reall, ``{General supersymmetric $AdS_5$ black
  holes},'' \href{http://dx.doi.org/10.1088/1126-6708/2004/04/048}{{\em JHEP}
  {\bfseries 0404} (2004) 048},
\href{http://arxiv.org/abs/hep-th/0401129}{{\ttfamily arXiv:hep-th/0401129
  [hep-th]}}.

\bibitem{Gauntlett:2004qy}
J.~P. Gauntlett and J.~B. Gutowski, ``{General Concentric Black Rings},''
  \href{http://dx.doi.org/10.1103/PhysRevD.71.045002}{{\em Phys.Rev.}
  {\bfseries D71} (2005) 045002},
\href{http://arxiv.org/abs/hep-th/0408122}{{\ttfamily arXiv:hep-th/0408122
  [hep-th]}}.

\bibitem{Gutowski:2005id}
J.~B. Gutowski and W.~Sabra, ``{General Supersymmetric Solutions of
  Five-Dimensional Supergravity},''
  \href{http://dx.doi.org/10.1088/1126-6708/2005/10/039}{{\em JHEP} {\bfseries
  0510} (2005) 039},
\href{http://arxiv.org/abs/hep-th/0505185}{{\ttfamily arXiv:hep-th/0505185
  [hep-th]}}.

\bibitem{Bellorin:2006yr}
J.~Bellorin, P.~Meessen, and T.~Ortin, ``{All the supersymmetric solutions of
  $N=1, d=5$ ungauged supergravity},''
  \href{http://dx.doi.org/10.1088/1126-6708/2007/01/020}{{\em JHEP} {\bfseries
  0701} (2007) 020},
\href{http://arxiv.org/abs/hep-th/0610196}{{\ttfamily arXiv:hep-th/0610196
  [hep-th]}}.

\bibitem{Bellorin:2007yp}
J.~Bellorin and T.~Ortin, ``{Characterization of all the supersymmetric
  solutions of gauged $N=1, d=5$ supergravity},''
  \href{http://dx.doi.org/10.1088/1126-6708/2007/08/096}{{\em JHEP} {\bfseries
  0708} (2007) 096},
\href{http://arxiv.org/abs/0705.2567}{{\ttfamily arXiv:0705.2567 [hep-th]}}.

\bibitem{Bellorin:2008we}
J.~Bellorin, ``{Supersymmetric solutions of gauged five-dimensional
  supergravity with general matter couplings},''
  \href{http://dx.doi.org/10.1088/0264-9381/26/19/195012}{{\em
  Class.Quant.Grav.} {\bfseries 26} (2009) 195012},
\href{http://arxiv.org/abs/0810.0527}{{\ttfamily arXiv:0810.0527 [hep-th]}}.

\bibitem{Bates:2003vx}
B.~Bates and F.~Denef, ``{Exact solutions for supersymmetric stationary black
  hole composites},'' \href{http://dx.doi.org/10.1007/JHEP11(2011)127}{{\em
  JHEP} {\bfseries 1111} (2011) 127},
\href{http://arxiv.org/abs/hep-th/0304094}{{\ttfamily arXiv:hep-th/0304094
  [hep-th]}}.

\bibitem{Ferrara:1995ih}
S.~Ferrara, R.~Kallosh, and A.~Strominger, ``{N=2 Extremal Black Holes},''
  \href{http://dx.doi.org/10.1103/PhysRevD.52.R5412}{{\em Phys.Rev.} {\bfseries
  D52} (1995) 5412--5416},
\href{http://arxiv.org/abs/hep-th/9508072}{{\ttfamily arXiv:hep-th/9508072
  [hep-th]}}.

\bibitem{Strominger:1996kf}
A.~Strominger, ``{Macroscopic Entropy of N=2 Extremal Black Holes},''
  \href{http://dx.doi.org/10.1016/0370-2693(96)00711-3}{{\em Phys.Lett.}
  {\bfseries B383} (1996) 39--43},
\href{http://arxiv.org/abs/hep-th/9602111}{{\ttfamily arXiv:hep-th/9602111
  [hep-th]}}.

\bibitem{Ferrara:1996dd}
S.~Ferrara and R.~Kallosh, ``{Supersymmetry and attractors},''
  \href{http://dx.doi.org/10.1103/PhysRevD.54.1514}{{\em Phys.Rev.} {\bfseries
  D54} (1996) 1514--1524},
\href{http://arxiv.org/abs/hep-th/9602136}{{\ttfamily arXiv:hep-th/9602136
  [hep-th]}}.

\bibitem{Ferrara:1996um}
S.~Ferrara and R.~Kallosh, ``{Universality of supersymmetric attractors},''
  \href{http://dx.doi.org/10.1103/PhysRevD.54.1525}{{\em Phys.Rev.} {\bfseries
  D54} (1996) 1525--1534},
\href{http://arxiv.org/abs/hep-th/9603090}{{\ttfamily arXiv:hep-th/9603090
  [hep-th]}}.

\bibitem{Moore:2004fg}
G.~W. Moore, ``{Les Houches lectures on strings and arithmetic},''
\href{http://arxiv.org/abs/hep-th/0401049}{{\ttfamily arXiv:hep-th/0401049
  [hep-th]}}.

\bibitem{Kraus:2005gh}
P.~Kraus and F.~Larsen, ``{Attractors and black rings},''
  \href{http://dx.doi.org/10.1103/PhysRevD.72.024010}{{\em Phys.Rev.}
  {\bfseries D72} (2005) 024010},
\href{http://arxiv.org/abs/hep-th/0503219}{{\ttfamily arXiv:hep-th/0503219
  [hep-th]}}.

\bibitem{Larsen:2006xm}
F.~Larsen, ``{The Attractor Mechanism in Five Dimensions},'' {\em Lect.Notes
  Phys.} {\bfseries 755} (2008) 249--281,
\href{http://arxiv.org/abs/hep-th/0608191}{{\ttfamily arXiv:hep-th/0608191
  [hep-th]}}.

\bibitem{Denef:2000ar}
F.~Denef, ``{On the correspondence between D-branes and stationary supergravity
  solutions of type II Calabi-Yau compactifications},''
\href{http://arxiv.org/abs/hep-th/0010222}{{\ttfamily arXiv:hep-th/0010222
  [hep-th]}}.

\bibitem{Denef:2001ix}
F.~Denef, ``{(Dis)assembling special Lagrangians},''
\href{http://arxiv.org/abs/hep-th/0107152}{{\ttfamily arXiv:hep-th/0107152
  [hep-th]}}.

\bibitem{Moore2010pitp}
G.~W. Moore, ``{PiTP Lectures on BPS States and Wall-Crossing in $d=4,
  \mathcal{N}=2$ Theories},''.
  \url{http://www.sns.ias.edu/pitp2/2010files/Moore_LectureNotes.rev3.pdf}.

\bibitem{Denef:2007vg}
F.~Denef and G.~W. Moore, ``{Split States, Entropy Enigmas, Holes and Halos},''
  \href{http://dx.doi.org/10.1007/JHEP11(2011)129}{{\em JHEP} {\bfseries 1111}
  (2011) 129},
\href{http://arxiv.org/abs/hep-th/0702146}{{\ttfamily arXiv:hep-th/0702146
  [hep-th]}}.

\bibitem{Bena:2005va}
I.~Bena and N.~P. Warner, ``{Bubbling Supertubes and Foaming Black Holes},''
  \href{http://dx.doi.org/10.1103/PhysRevD.74.066001}{{\em Phys.Rev.}
  {\bfseries D74} (2006) 066001},
\href{http://arxiv.org/abs/hep-th/0505166}{{\ttfamily arXiv:hep-th/0505166
  [hep-th]}}.

\bibitem{Berglund:2005vb}
P.~Berglund, E.~G. Gimon, and T.~S. Levi, ``{Supergravity Microstates for BPS
  Black Holes and Black Rings},''
  \href{http://dx.doi.org/10.1088/1126-6708/2006/06/007}{{\em JHEP} {\bfseries
  0606} (2006) 007},
\href{http://arxiv.org/abs/hep-th/0505167}{{\ttfamily arXiv:hep-th/0505167
  [hep-th]}}.

\bibitem{Mathur:2005zp}
S.~D. Mathur, ``{The Fuzzball proposal for black holes: An Elementary
  review},'' \href{http://dx.doi.org/10.1002/prop.200410203}{{\em
  Fortsch.Phys.} {\bfseries 53} (2005) 793--827},
\href{http://arxiv.org/abs/hep-th/0502050}{{\ttfamily arXiv:hep-th/0502050
  [hep-th]}}.

\bibitem{Bergshoeff:2011se}
E.~A. Bergshoeff, T.~Ortin, and F.~Riccioni, ``{Defect Branes},''
  \href{http://dx.doi.org/10.1016/j.nuclphysb.2011.10.037}{{\em Nucl.Phys.}
  {\bfseries B856} (2012) 210--227},
\href{http://arxiv.org/abs/1109.4484}{{\ttfamily arXiv:1109.4484 [hep-th]}}.

\bibitem{Kikuchi:2012za}
T.~Kikuchi, T.~Okada, and Y.~Sakatani, ``{Rotating string in doubled geometry
  with generalized isometries},''
  \href{http://dx.doi.org/10.1103/PhysRevD.86.046001}{{\em Phys.Rev.}
  {\bfseries D86} (2012) 046001},
\href{http://arxiv.org/abs/1205.5549}{{\ttfamily arXiv:1205.5549 [hep-th]}}.

\bibitem{Andriot:2013xca}
D.~Andriot and A.~Betz, ``{$\beta$-supergravity: a ten-dimensional theory with
  non-geometric fluxes, and its geometric framework},''
  \href{http://dx.doi.org/10.1007/JHEP12(2013)083}{{\em JHEP} {\bfseries 1312}
  (2013) 083},
\href{http://arxiv.org/abs/1306.4381}{{\ttfamily arXiv:1306.4381 [hep-th]}}.

\bibitem{Geissbuhler:2013uka}
D.~Geissbuhler, D.~Marques, C.~Nunez, and V.~Penas, ``{Exploring Double Field
  Theory},'' \href{http://dx.doi.org/10.1007/JHEP06(2013)101}{{\em JHEP}
  {\bfseries 1306} (2013) 101},
\href{http://arxiv.org/abs/1304.1472}{{\ttfamily arXiv:1304.1472 [hep-th]}}.

\bibitem{Kimura:2013fda}
T.~Kimura and S.~Sasaki, ``{Gauged Linear Sigma Model for Exotic Five-brane},''
  \href{http://dx.doi.org/10.1016/j.nuclphysb.2013.08.017}{{\em Nucl.Phys.}
  {\bfseries B876} (2013) 493--508},
\href{http://arxiv.org/abs/1304.4061}{{\ttfamily arXiv:1304.4061 [hep-th]}}.

\bibitem{Hassler:2013wsa}
F.~Hassler and D.~Lust, ``{Non-commutative/non-associative IIA (IIB) Q- and
  R-branes and their intersections},''
  \href{http://dx.doi.org/10.1007/JHEP07(2013)048}{{\em JHEP} {\bfseries 1307}
  (2013) 048},
\href{http://arxiv.org/abs/1303.1413}{{\ttfamily arXiv:1303.1413 [hep-th]}}.

\bibitem{Kimura:2013zva}
T.~Kimura and S.~Sasaki, ``{Worldsheet instanton corrections to $5^2_2$-brane
  geometry},'' \href{http://dx.doi.org/10.1007/JHEP08(2013)126}{{\em JHEP}
  {\bfseries 1308} (2013) 126},
\href{http://arxiv.org/abs/1305.4439}{{\ttfamily arXiv:1305.4439 [hep-th]}}.

\bibitem{Chatzistavrakidis:2013jqa}
A.~Chatzistavrakidis, F.~F. Gautason, G.~Moutsopoulos, and M.~Zagermann,
  ``{Effective actions of nongeometric five-branes},''
  \href{http://dx.doi.org/10.1103/PhysRevD.89.066004}{{\em Phys.Rev.}
  {\bfseries D89} no.~6, (2014) 066004},
\href{http://arxiv.org/abs/1309.2653}{{\ttfamily arXiv:1309.2653 [hep-th]}}.

\bibitem{Andriot:2014uda}
D.~Andriot and A.~Betz, ``{NS-branes, source corrected Bianchi identities, and
  more on backgrounds with non-geometric fluxes},''
  \href{http://dx.doi.org/10.1007/JHEP07(2014)059}{{\em JHEP} {\bfseries 1407}
  (2014) 059},
\href{http://arxiv.org/abs/1402.5972}{{\ttfamily arXiv:1402.5972 [hep-th]}}.

\bibitem{Kimura:2013khz}
T.~Kimura and S.~Sasaki, ``{Worldsheet Description of Exotic Five-brane with
  Two Gauged Isometries},''
  \href{http://dx.doi.org/10.1007/JHEP03(2014)128}{{\em JHEP} {\bfseries 1403}
  (2014) 128},
\href{http://arxiv.org/abs/1310.6163}{{\ttfamily arXiv:1310.6163 [hep-th]}}.

\bibitem{Kimura:2014upa}
T.~Kimura, S.~Sasaki, and M.~Yata, ``{World-volume Effective Actions of Exotic
  Five-branes},'' \href{http://dx.doi.org/10.1007/JHEP07(2014)127}{{\em JHEP}
  {\bfseries 1407} (2014) 127},
\href{http://arxiv.org/abs/1404.5442}{{\ttfamily arXiv:1404.5442 [hep-th]}}.

\bibitem{Okada:2014wma}
T.~Okada and Y.~Sakatani, ``{Defect branes as Alice strings},''
  \href{http://dx.doi.org/10.1007/JHEP03(2015)131}{{\em JHEP} {\bfseries 1503}
  (2015) 131},
\href{http://arxiv.org/abs/1411.1043}{{\ttfamily arXiv:1411.1043 [hep-th]}}.

\bibitem{Kimura:2014wga}
T.~Kimura, ``{Defect $(p,q)$ Five-branes},''
  \href{http://dx.doi.org/10.1016/j.nuclphysb.2015.01.023}{{\em Nucl.Phys.}
  {\bfseries B893} (2015) 1--20},
\href{http://arxiv.org/abs/1410.8403}{{\ttfamily arXiv:1410.8403 [hep-th]}}.

\bibitem{Sakatani:2014hba}
Y.~Sakatani, ``{Exotic branes and non-geometric fluxes},''
  \href{http://dx.doi.org/10.1007/JHEP03(2015)135}{{\em JHEP} {\bfseries 03}
  (2015) 135},
\href{http://arxiv.org/abs/1412.8769}{{\ttfamily arXiv:1412.8769 [hep-th]}}.

\bibitem{Kimura:2014bea}
T.~Kimura, S.~Sasaki, and M.~Yata, ``{Hyper-K{\"a}hler with torsion, T-duality,
  and defect (p, q) five-branes},''
  \href{http://dx.doi.org/10.1007/JHEP03(2015)076}{{\em JHEP} {\bfseries 1503}
  (2015) 076},
\href{http://arxiv.org/abs/1411.3457}{{\ttfamily arXiv:1411.3457 [hep-th]}}.

\bibitem{Kimura:2015yla}
T.~Kimura, ``{$\mathcal{N}=(4,4)$ Gauged Linear Sigma Models for Defect
  Five-branes},''
\href{http://arxiv.org/abs/1503.08635}{{\ttfamily arXiv:1503.08635 [hep-th]}}.

\bibitem{Gibbons:1987sp}
G.~Gibbons and P.~Ruback, ``{The Hidden Symmetries of Multicenter Metrics},''
\href{http://dx.doi.org/10.1007/BF01466773}{{\em Commun.Math.Phys.} {\bfseries
  115} (1988) 267}.

\bibitem{Gibbons:1979zt}
G.~Gibbons and S.~Hawking, ``{Gravitational Multi-Instantons},''
\href{http://dx.doi.org/10.1016/0370-2693(78)90478-1}{{\em Phys.Lett.}
  {\bfseries B78} (1978) 430}.

\bibitem{Strominger:1996sh}
A.~Strominger and C.~Vafa, ``{Microscopic Origin of the Bekenstein-Hawking
  Entropy},'' \href{http://dx.doi.org/10.1016/0370-2693(96)00345-0}{{\em
  Phys.Lett.} {\bfseries B379} (1996) 99--104},
\href{http://arxiv.org/abs/hep-th/9601029}{{\ttfamily arXiv:hep-th/9601029
  [hep-th]}}.

\bibitem{Breckenridge:1996is}
J.~Breckenridge, R.~C. Myers, A.~Peet, and C.~Vafa, ``{D-branes and spinning
  black holes},'' \href{http://dx.doi.org/10.1016/S0370-2693(96)01460-8}{{\em
  Phys.Lett.} {\bfseries B391} (1997) 93--98},
\href{http://arxiv.org/abs/hep-th/9602065}{{\ttfamily arXiv:hep-th/9602065
  [hep-th]}}.

\bibitem{Elvang:2004rt}
H.~Elvang, R.~Emparan, D.~Mateos, and H.~S. Reall, ``{A Supersymmetric black
  ring},'' \href{http://dx.doi.org/10.1103/PhysRevLett.93.211302}{{\em
  Phys.Rev.Lett.} {\bfseries 93} (2004) 211302},
\href{http://arxiv.org/abs/hep-th/0407065}{{\ttfamily arXiv:hep-th/0407065
  [hep-th]}}.

\bibitem{Elvang:2004ds}
H.~Elvang, R.~Emparan, D.~Mateos, and H.~S. Reall, ``{Supersymmetric black
  rings and three-charge supertubes},''
  \href{http://dx.doi.org/10.1103/PhysRevD.71.024033}{{\em Phys.Rev.}
  {\bfseries D71} (2005) 024033},
\href{http://arxiv.org/abs/hep-th/0408120}{{\ttfamily arXiv:hep-th/0408120
  [hep-th]}}.

\bibitem{Maldacena:1997de}
J.~M. Maldacena, A.~Strominger, and E.~Witten, ``{Black Hole Entropy in
  M-theory},'' \href{http://dx.doi.org/10.1088/1126-6708/1997/12/002}{{\em
  JHEP} {\bfseries 9712} (1997) 002},
\href{http://arxiv.org/abs/hep-th/9711053}{{\ttfamily arXiv:hep-th/9711053
  [hep-th]}}.

\bibitem{Emparan:2001ux}
R.~Emparan, D.~Mateos, and P.~K. Townsend, ``{Supergravity Supertubes},''
  \href{http://dx.doi.org/10.1088/1126-6708/2001/07/011}{{\em JHEP} {\bfseries
  0107} (2001) 011},
\href{http://arxiv.org/abs/hep-th/0106012}{{\ttfamily arXiv:hep-th/0106012
  [hep-th]}}.

\bibitem{Lunin:2001fv}
O.~Lunin and S.~D. Mathur, ``{Metric of the multiply wound rotating string},''
  \href{http://dx.doi.org/10.1016/S0550-3213(01)00321-2}{{\em Nucl.Phys.}
  {\bfseries B610} (2001) 49--76},
\href{http://arxiv.org/abs/hep-th/0105136}{{\ttfamily arXiv:hep-th/0105136
  [hep-th]}}.

\bibitem{Niehoff:2012wu}
B.~E. Niehoff, O.~Vasilakis, and N.~P. Warner, ``{Multi-Superthreads and
  Supersheets},'' \href{http://dx.doi.org/10.1007/JHEP04(2013)046}{{\em JHEP}
  {\bfseries 1304} (2013) 046},
\href{http://arxiv.org/abs/1203.1348}{{\ttfamily arXiv:1203.1348 [hep-th]}}.

\bibitem{Bena:2011dd}
I.~Bena, S.~Giusto, M.~Shigemori, and N.~P. Warner, ``{Supersymmetric Solutions
  in Six Dimensions: A Linear Structure},''
  \href{http://dx.doi.org/10.1007/JHEP03(2012)084}{{\em JHEP} {\bfseries 1203}
  (2012) 084},
\href{http://arxiv.org/abs/1110.2781}{{\ttfamily arXiv:1110.2781 [hep-th]}}.

\bibitem{Marolf:2000cb}
D.~Marolf, ``{Chern-Simons Terms and the Three Notions of Charge},''
\href{http://arxiv.org/abs/hep-th/0006117}{{\ttfamily arXiv:hep-th/0006117
  [hep-th]}}.

\bibitem{Arnowitt:1962hi}
R.~L. Arnowitt, S.~Deser, and C.~W. Misner, ``{The Dynamics of General
  Relativity},'' \href{http://dx.doi.org/10.1007/s10714-008-0661-1}{{\em
  Gen.Rel.Grav.} {\bfseries 40} (2008) 1997--2027},
\href{http://arxiv.org/abs/gr-qc/0405109}{{\ttfamily arXiv:gr-qc/0405109
  [gr-qc]}}.

\bibitem{Bena:2007kg}
I.~Bena and N.~P. Warner, ``{Black holes, black rings and their microstates},''
  \href{http://dx.doi.org/10.1007/978-3-540-79523-0_1}{{\em Lect.Notes Phys.}
  {\bfseries 755} (2008) 1--92},
\href{http://arxiv.org/abs/hep-th/0701216}{{\ttfamily arXiv:hep-th/0701216
  [hep-th]}}.

\bibitem{Skenderis:2008qn}
K.~Skenderis and M.~Taylor, ``{The fuzzball proposal for black holes},''
  \href{http://dx.doi.org/10.1016/j.physrep.2008.08.001}{{\em Phys.Rept.}
  {\bfseries 467} (2008) 117--171},
\href{http://arxiv.org/abs/0804.0552}{{\ttfamily arXiv:0804.0552 [hep-th]}}.

\bibitem{Balasubramanian:2008da}
V.~Balasubramanian, J.~de~Boer, S.~El-Showk, and I.~Messamah, ``{Black Holes as
  Effective Geometries},''
  \href{http://dx.doi.org/10.1088/0264-9381/25/21/214004}{{\em
  Class.Quant.Grav.} {\bfseries 25} (2008) 214004},
\href{http://arxiv.org/abs/0811.0263}{{\ttfamily arXiv:0811.0263 [hep-th]}}.

\bibitem{Chowdhury:2010ct}
B.~D. Chowdhury and A.~Virmani, ``{Modave Lectures on Fuzzballs and Emission
  from the D1-D5 System},''
\href{http://arxiv.org/abs/1001.1444}{{\ttfamily arXiv:1001.1444 [hep-th]}}.

\bibitem{deBoer:2009un}
J.~de~Boer, S.~El-Showk, I.~Messamah, and D.~Van~den Bleeken, ``{A Bound on the
  Entropy of Supergravity?},''
  \href{http://dx.doi.org/10.1007/JHEP02(2010)062}{{\em JHEP} {\bfseries 1002}
  (2010) 062},
\href{http://arxiv.org/abs/0906.0011}{{\ttfamily arXiv:0906.0011 [hep-th]}}.

\bibitem{Bena:2010gg}
I.~Bena, N.~Bobev, S.~Giusto, C.~Ruef, and N.~P. Warner, ``{An
  Infinite-Dimensional Family of Black-Hole Microstate Geometries},''
  \href{http://dx.doi.org/10.1007/JHEP03(2011)022,
  10.1007/JHEP04(2011)059}{{\em JHEP} {\bfseries 1103} (2011) 022},
\href{http://arxiv.org/abs/1006.3497}{{\ttfamily arXiv:1006.3497 [hep-th]}}.

\bibitem{Bena:2006kb}
I.~Bena, C.-W. Wang, and N.~P. Warner, ``{Mergers and Typical Black Hole
  Microstates},'' \href{http://dx.doi.org/10.1088/1126-6708/2006/11/042}{{\em
  JHEP} {\bfseries 0611} (2006) 042},
\href{http://arxiv.org/abs/hep-th/0608217}{{\ttfamily arXiv:hep-th/0608217
  [hep-th]}}.

\bibitem{Bena:2007qc}
I.~Bena, C.-W. Wang, and N.~P. Warner, ``{Plumbing the Abyss: Black ring
  microstates},'' \href{http://dx.doi.org/10.1088/1126-6708/2008/07/019}{{\em
  JHEP} {\bfseries 0807} (2008) 019},
\href{http://arxiv.org/abs/0706.3786}{{\ttfamily arXiv:0706.3786 [hep-th]}}.

\bibitem{Bena:2012hf}
I.~Bena, M.~Berkooz, J.~de~Boer, S.~El-Showk, and D.~Van~den Bleeken,
  ``{Scaling BPS Solutions and pure-Higgs States},''
  \href{http://dx.doi.org/10.1007/JHEP11(2012)171}{{\em JHEP} {\bfseries 1211}
  (2012) 171},
\href{http://arxiv.org/abs/1205.5023}{{\ttfamily arXiv:1205.5023 [hep-th]}}.

\bibitem{Lee:2012sc}
S.-J. Lee, Z.-L. Wang, and P.~Yi, ``{Quiver Invariants from Intrinsic Higgs
  States},'' \href{http://dx.doi.org/10.1007/JHEP07(2012)169}{{\em JHEP}
  {\bfseries 1207} (2012) 169},
\href{http://arxiv.org/abs/1205.6511}{{\ttfamily arXiv:1205.6511 [hep-th]}}.

\bibitem{Bena:2011uw}
I.~Bena, J.~de~Boer, M.~Shigemori, and N.~P. Warner, ``{Double, Double
  Supertube Bubble},'' \href{http://dx.doi.org/10.1007/JHEP10(2011)116}{{\em
  JHEP} {\bfseries 1110} (2011) 116},
\href{http://arxiv.org/abs/1107.2650}{{\ttfamily arXiv:1107.2650 [hep-th]}}.

\bibitem{Bena:2014qxa}
I.~Bena, M.~Shigemori, and N.~P. Warner, ``{Black-Hole Entropy from
  Supergravity Superstrata States},''
  \href{http://dx.doi.org/10.1007/JHEP10(2014)140}{{\em JHEP} {\bfseries 1410}
  (2014) 140},
\href{http://arxiv.org/abs/1406.4506}{{\ttfamily arXiv:1406.4506 [hep-th]}}.

\bibitem{Bergshoeff:2006jj}
E.~A. Bergshoeff, J.~Hartong, T.~Ortin, and D.~Roest, ``{Seven-branes and
  Supersymmetry},'' \href{http://dx.doi.org/10.1088/1126-6708/2007/02/003}{{\em
  JHEP} {\bfseries 0702} (2007) 003},
\href{http://arxiv.org/abs/hep-th/0612072}{{\ttfamily arXiv:hep-th/0612072
  [hep-th]}}.

\bibitem{Niehoff:2013kia}
B.~E. Niehoff and N.~P. Warner, ``{Doubly-Fluctuating BPS Solutions in Six
  Dimensions},'' \href{http://dx.doi.org/10.1007/JHEP10(2013)137}{{\em JHEP}
  {\bfseries 1310} (2013) 137},
\href{http://arxiv.org/abs/1303.5449}{{\ttfamily arXiv:1303.5449 [hep-th]}}.

\end{thebibliography}

\providecommand{\href}[2]{#2}\begingroup\raggedright\endgroup


\end{document}